\documentclass[aps,physrev,preprint,groupedaddress]{revtex4-2}

\usepackage{mathtools}
\usepackage{color}
\usepackage{subfigure}
\usepackage[T1]{fontenc}

\begin{document}

\title{Interpretation of Unfair Sampling in Quantum Annealing \\by Node Centrality}

\author{Naoki Maruyama}
\email{Contact author: naoki.maruyama.p7@dc.tohoku.ac.jp}
\affiliation{Graduate School of Information Science, Tohoku University, Sendai 980-8579, Japan}
\affiliation{Sigma-i Co., Ltd., Tokyo 108-0075, Japan}

\author{Masayuki Ohzeki}
\affiliation{Graduate School of Information Science, Tohoku University, Sendai 980-8579, Japan}
\affiliation{Department of Physics, Institute of Science Tokyo, Tokyo 152-8551, Japan}
\affiliation{Research and Education Institute for Semiconductors and Informatics, Kumamoto University, Kumamoto 860-8555, Japan}
\affiliation{Sigma-i Co., Ltd., Tokyo 108-0075, Japan}

\begin{abstract}
In applications where multiple optimal solutions are needed, transverse-field quantum annealing (QA) is known to sample degenerate ground states in a strongly biased manner.
Despite extensive empirical observations, it remains unclear which features of degenerate ground states are preferentially sampled and why by QA.
Here we analyze the final states using degenerate perturbation theory to characterize the preference among them.
In this analysis, the adjacency matrix of the graph composed by the ground states naturally emerges, and we can predict the eigenvector centralities (one of the node centralities) are related to the probabilities of these states.
We verify this prediction on toy models where degeneracy is lifted at first and second order, and we show that second-order weights encode local barrier information, relating sampling fairness to the flatness of the local energy landscape.
Finally, this perspective suggests two practical routes toward fair sampling --- promoting connectivity of the graph and reducing heterogeneity of centralities --- and we illustrate consistency with higher-order drivers and minor-embedding transformations.
\end{abstract}

\maketitle

\section{Introduction}

Quantum annealing (QA) is an algorithm that searches for optimal solutions to combinatorial optimization problems by exploiting quantum effects \cite{kadowaki1998, farhi2001}.
D-Wave Systems has commercialized hardware that physically implements QA, and improvements such as increasing the number of qubits and reducing errors have been made so far \cite{johnson2010, dattani2019, boothby2020}.
The applicability of QA to optimization problems in various domains has been investigated, including traffic control \cite{neukart2017, shikanai2025}, production scheduling \cite{venturelli2016, sawamura2025}, and logistics \cite{feld2019, haba2025}.
Furthermore, the effectiveness of QA has been demonstrated in various areas, including quantum chemistry calculations \cite{streif2019, genin2019}, quantum simulation \cite{king2022}, and machine learning \cite{neven2012, sato2021}.

Some real optimization problems require not a single optimal solution but diverse near-optimal solutions.
For example, in chemical material discovery, it is necessary not only to find molecules with desirable property values but also to consider compositional feasibility.
Since specific structures can be difficult to synthesize or fabricate, it is often essential to propose diverse candidates.
Moreover, beyond specific domains, diverse solutions prove valuable for enhancing search efficiency and improving solution quality across various optimization methods, including genetic programming, robust optimization, and multi-objective optimization.
It has been reported that QA can obtain diverse samples compared with classical sampling methods \cite{zucca2021, mohseni2021}.
QA has recently been utilized for black-box optimization that requires such solutions \cite{baptista2018, koshikawa2021}, and this property is believed to underlie its success in this context.
Black-box optimization using QA has been applied to materials discovery in practice \cite{kitai2020, doi2023}, and the numerical results demonstrate that QA-based optimization can yield more diverse solutions than existing methods.
As another example, the usefulness of diverse samples has been reported in tasks that remove incorrectly labeled instances from contaminated training datasets \cite{otsuka2025a}.

While the diversity of approximate solutions obtained by QA has attracted attention, if we focus on only optimal solutions, standard QA fails to obtain multiple optimal solutions uniformly \cite{matsuda2009}.
This phenomenon is known as unfair sampling and has also been observed in experiments using the quantum annealer \cite{mandra2017, pelofske2023}.
There are many applications in which one desires to obtain multiple optimal solutions equally, such as SAT filters \cite{weaver2012, azinovic2017} and machine learning \cite{dixit2021a, sato2021}.
Unfair sampling limits the applicability of QA to problems that require diverse optimal solutions.

Although methods for mitigating unfairness in QA have been studied, a standard approach has yet to be established.
It is known that higher-order fully connected drivers can achieve uniform sampling among degenerate ground states by uniformly coupling ground states \cite{matsuda2009, konz2019}, but implementing such complex drivers in hardware is currently challenging.
The current quantum annealer has a function to incorporate thermal effects depending on the choice of annealing schedule, which can reduce the bias of unfairness \cite{kadowaki2019}.
In the method proposed by the previous study \cite{kumar2020}, random perturbations lift the degeneracy of ground states, allowing a different ground state to be selected in each run, and overall, the sampling becomes nearly uniform.
On the other hand, while QA with a transverse field exhibits unfairness, classical methods such as simulated annealing (SA) can comparatively easily achieve fair sampling.
This is because, by evolving a Markov chain Monte Carlo process adiabatically in time, the system reaches thermal equilibrium at each time, and the Boltzmann distribution at thermal equilibrium assigns equal probabilities to states with equal energies \cite{geman1984}.
For example, the Tempering-based classical algorithm \cite{chancellor2017, zhu2019} and the method to perform SA on QA \cite{somma2007, yamamoto2020} have been proposed.

Understanding the mechanism of unfair sampling in QA is crucial for developing methods that ensure fair sampling, as no established method currently exists.
However, it is still unclear what characteristics of ground states QA tends to favor.
One relevant study \cite{matsuda2009} reports that spins capable of flipping without energy change are called "free spins", and that free spins are sampled more frequently.
In this study, we clarify the preference over ground states using perturbation theory.
In perturbation theory for a degenerate system, an adjacency matrix over the ground states naturally emerges, and we focus on the fact that eigenvector centrality --- one of several node-centrality measures --- in the graph composed of these ground states relates to their probabilities.
Our experiments on toy models show that ground states with higher eigenvector centrality tend to have higher probabilities.
When the ground states are significantly separated for a given driver, we find that the flatness of the energy landscape around the ground state is related to the fairness of solutions.
Furthermore, based on the above interpretation via node centrality, we organize guidelines for achieving fair sampling.
As examples that fit these guidelines, we confirm that using higher-order drivers and minor embedding onto a hardware graph can interpret the fairness in terms of centrality.

The remainder of this paper is organized as follows.
In Sec. II, we review QA and introduce solution graphs arising from degenerate perturbation theory, together with a centrality-based interpretation.
In Sec. III, we present numerical results for first- and second-order perturbations and discuss guidelines for fair sampling, connecting them to higher-order drivers and minor embedding.
Section IV concludes with a summary and outlook.

\section{Methods}

Quantum annealing (QA) is a method that utilizes quantum effects to solve combinatorial optimization problems.
A combinatorial optimization problem is equivalent to an Ising model, and the following Hamiltonian represents its cost function.
\begin{equation}
H_0 = -\sum_{i=1}^N h_i \sigma_i -\sum_{i=1}^{N} \sum_{j>i}^{N} J_{ij} \sigma_i \sigma_j,
\end{equation}
where $\sigma_i \in \{\pm1\}$ is the $i$-th Ising variable, $h_i$ is a local field, $J_{ij}$ is the interaction coefficient between spins, and $N$ is the number of spins.
In QA, the Hamiltonian of the quantized Ising model is given by
\begin{equation}
H(t)=\frac{t}{\tau} H_0\left(\boldsymbol{\sigma}^z\right) + \left(1-\frac{t}{\tau}\right) V \left( \boldsymbol{\sigma}^x \right),
\end{equation}
where $\boldsymbol{\sigma}^{x}$ and $\boldsymbol{\sigma}^{z}$ are the $x,z$-components of the Pauli matrices of spins, and $\tau$ is the annealing time.
The first term $H_0$ is called the target Hamiltonian and represents the original problem.
The second term $V$ is referred to as the driver Hamiltonian and represents the quantum fluctuations that drive the search for solutions.
Except for some experiments, we use the transverse-field driver Hamiltonian $V = -\sum_{i=1}^N \sigma_i^x$ in this study.
In standard QA, the system evolves according to the Schrödinger equation.
In the QA process, we prepare at the initial time $t=0$ a state that is an equal superposition of all solutions and slowly weaken the quantum fluctuations until $t = \tau$.
If this time variation is sufficiently slow, the adiabatic condition ensures that the final state becomes the ground state \cite{morita2008}.

Our study investigates the types of states favored by analyzing the final states obtained analytically using perturbation theory, as in previous works \cite{sieberer2018, konz2019}.
We consider the case where the system evolves in time according to the Schrödinger equation with $\tau$ sufficiently large, and the quantum fluctuations are varied slowly enough in time to satisfy the adiabatic condition.
In this case, the instantaneous excited states are exponentially suppressed, which means towards the end of annealing at $\tau - \lambda$ (for a small $\lambda > 0$) the system is in the ground state of $H\left( \tau - \lambda \right)$.
Because $H\left(\tau - \lambda\right)$ can be viewed as $H_{0}$ perturbed by $V$, we analyze the probabilities of the ground states using a perturbative approach.
Then, we define the Hamiltonian around the final time as $H(\lambda) = H_{0} + \lambda V$,
where $\lambda > 0$ is a sufficiently small coefficient.
The eigenstate $|n(\lambda)\rangle$ and eigenenergy $E_{n}^{\lambda}$ of $H(\lambda)$ are expanded as $|n(\lambda)\rangle=\left|n^{(0)}\right\rangle+\lambda\left|n^{(1)}\right\rangle+\lambda^{2}\left|n^{(2)}\right\rangle+\ldots$, $E_{n}(\lambda)=E_{n}^{(0)}+\lambda E_{n}^{(1)}+\lambda^{2} E_{n}^{(2)}+\ldots$ for small $\lambda$.

When the first-order perturbation lifts the degeneracy, the ground state $|n^{(0)}\rangle$ corresponding to the smallest eigenvalue $E_0^{(1)}$ is obtained as the final state by solving the following eigenvalue equation.
\begin{equation}
E_{n}^{(1)}|n^{(0)}\rangle=P_{1} V P_{1} |n^{(0)}\rangle,
\label{eq:1st_perturb_eigeneq}
\end{equation}
where $P_{1}$ is the projection operator onto the degenerate ground-state subspace of $H_{0}$.

\begin{figure}[tbh]
\begin{center}
\includegraphics[width=0.5\linewidth]{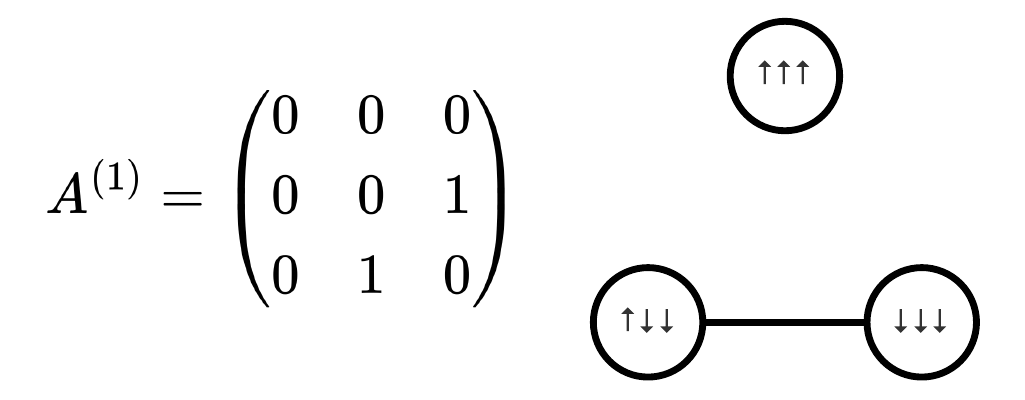}
\caption{
Example of adjacency matrix (left) and solution graph (right) in a model with degenerate ground states: $|\uparrow \uparrow \uparrow \rangle, |\uparrow \downarrow \downarrow \rangle, |\downarrow \downarrow \downarrow \rangle$, using a transverse-field driver.
}
\label{fig:ex_solution_graph}
\end{center}
\end{figure}

In first-order perturbation, $-P_1 V P_1 =: A^{(1)}$ can be interpreted as an "adjacency matrix" over the ground states.
Specifically, the matrix element $A_{ij}^{(1)}$ is unity if a given driver allows a direct transition from ground state $i$ to $j$, and zero otherwise.
For a transverse-field driver, elements between ground states that are connected by a single spin flip take the value one.
As an example, the left side of Fig.~\ref{fig:ex_solution_graph} shows the adjacency matrix for a model with three degenerate ground states $|\uparrow \uparrow \uparrow \rangle, |\uparrow \downarrow \downarrow \rangle, |\downarrow \downarrow \downarrow \rangle$ in a transverse-field driver.
The states $|\uparrow \downarrow \downarrow \rangle$ and $|\downarrow \downarrow \downarrow \rangle$ can transition to each other by a single spin flip.

In this work, we refer to the graph corresponding to the above adjacency matrix as a "solution graph".
This is a graph whose nodes are ground states and whose edges connect pairs of ground states that are mutually reachable via the driver.
The right side of Fig.~\ref{fig:ex_solution_graph} shows the solution graph corresponding to the adjacency matrix, where the states $|\uparrow \downarrow \downarrow \rangle$ and $|\downarrow \downarrow \downarrow \rangle$ that are connected by a single spin flip are joined by an edge.

An adjacency matrix contains almost all the information about the graph, and analyzing it can reveal its properties.
We focus here on node centrality and examine the preferences associated with each node (ground state).
Node centrality measures how central each node is within a graph.
Degree centrality is one of the simplest definitions, which assigns higher scores to nodes that have more connections (higher degree) in the graph.
If we denote an adjacency matrix by $A_{ij}$, the degree centrality of node $i$ is given by $c_i^{\mathrm{deg}} = \sum_{j=1}^N  A_{ij}$.
Whereas degree centrality weights all neighboring nodes equally, eigenvector centrality weights them by the centralities of the neighboring nodes themselves:
\begin{equation}
c_i^{\mathrm{eig}} = \frac{1}{\mu}\sum_{j=1}^{N}A_{ij} c_j^{\mathrm{eig}},
\label{eq:eigenvec_centrality}
\end{equation}
where $\mu$ is an eigenvalue.
Eigenvector centrality indirectly considers nodes that are not directly adjacent, providing an index that reflects the broader structure of the graph.

Eq.~\eqref{eq:1st_perturb_eigeneq} can be rewritten as $\mu_n\left|n^{(0)}\right\rangle=A^{(1)}\left|n^{(0)}\right\rangle$, where $\mu_n:=-E_n^{(1)}$, corresponding to Eq.~\eqref{eq:eigenvec_centrality}.
Therefore, the state selected within the degenerate manifold at first order is the eigenvector corresponding to the largest eigenvalue $\mu_1$ of $A^{(1)}$ (equivalently, the smallest $E^{(1)}$).
Let $\left\{\left|g_i\right\rangle\right\}_{i=1}^M$ denote the degenerate ground states $H_0$ (computational basis states).
The principal eigenvector of $A^{(1)}$ can be written as $\left|\psi^{(0)}\right\rangle=\sum_i c_i\left|g_i\right\rangle$ with $c_i \geq 0$ for connected graphs.
Upon measuring in the computational basis, the sampling probabilities satisfy $p_i \propto\left|c_i\right|^2$.
Hence, eigenvector centrality (up to normalization) provides a direct predictor of the relative sampling probabilities among degenerate ground states.

\section{Results}

\begin{figure}[tbh]
\centering
\subfigure[$N$-spin chain model]{\includegraphics[width=0.5\linewidth]{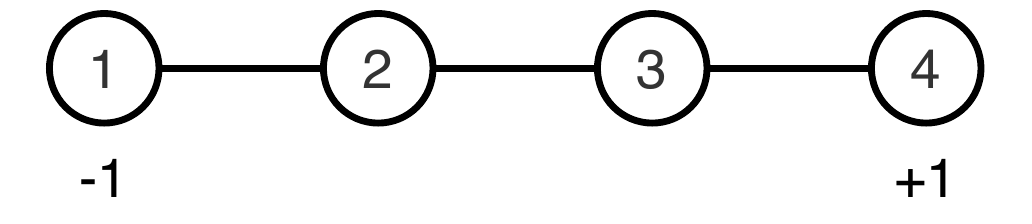}\label{fig:nspins_chain_model}}
\par\vspace{1em}
\subfigure[Solution graph]{\includegraphics[width=0.5\linewidth]{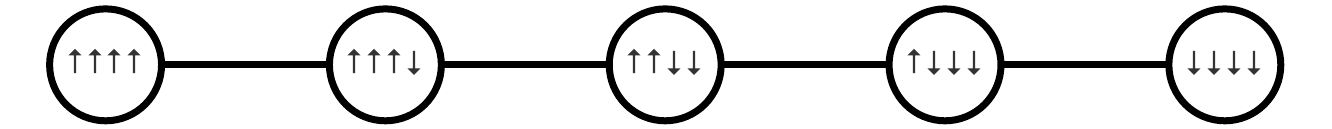}\label{fig:nspins_chain_solution_graph}}
\par\vspace{1em}
\subfigure[Eigenvector centrality and probability]{\includegraphics[width=0.6\linewidth]{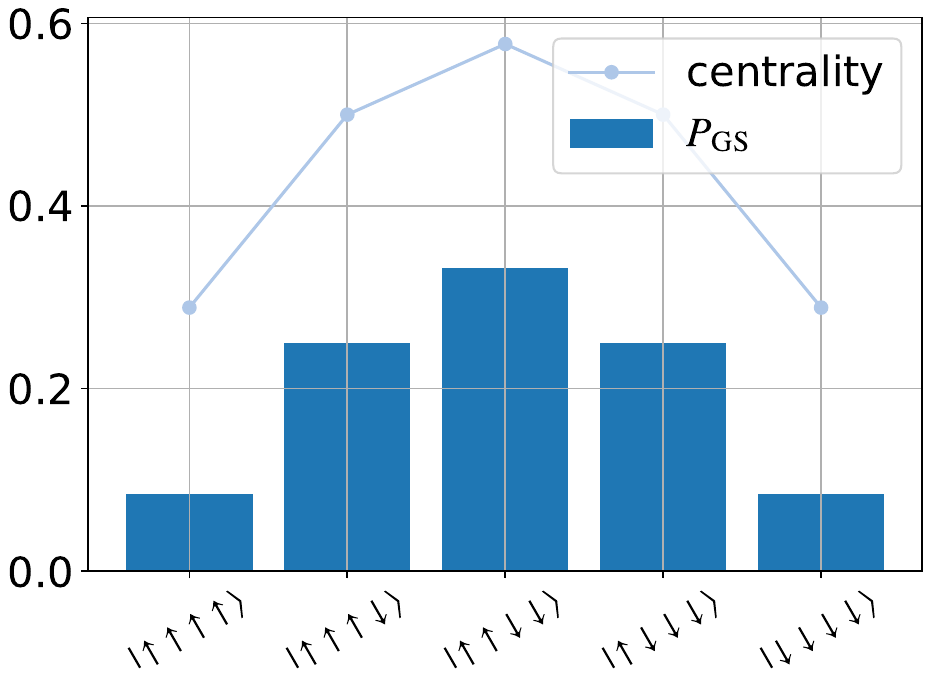}\label{fig:nspins_chain_results}}
\caption{
(a) $N$-spin chain model in $N=4$.
(b) Solution graph using a transverse-field driver.
(c) Ground-state probabilities $P_{\mathrm{GS}}$ (bars) compared with the eigenvector-centrality (line).
}
\label{fig:nspins_chain}
\end{figure}

We investigate whether the above prediction, relating eigenvector centrality and probability, holds in models where degeneracy is lifted at the first order.
The toy model considered here is $N$-spin chain model with opposite boundary fields (Fig.~\ref{fig:nspins_chain_model}), whose Hamiltonian is given by $H_0=-\sum_{i=1}^{N-1} \sigma_i^z \sigma_{i+1}^z-\sigma_1^z+\sigma_N^z$.
This model has $N+1$ ground states: $|\uparrow^{N-n} \downarrow^n\rangle \ (n=0, \ldots, N)$.
In a transverse field, the solution graph of this model ($N=4$) becomes a chain, as shown in Fig.~\ref{fig:nspins_chain_solution_graph}, and the adjacency matrix is given as follows:
\begin{equation}
A^{(1)} = \begin{pmatrix}
0 & 1 & 0 & 0 & 0 \\
1 & 0 & 1 & 0 & 0 \\
0 & 1 & 0 & 1 & 0 \\
0 & 0 & 1 & 0 & 1 \\
0 & 0 & 0 & 1 & 0
\end{pmatrix}.
\end{equation}

By computing the eigenvector centrality on this solution graph and comparing it with the probabilities obtained from perturbation theory, we find that the two distributions agree qualitatively, as shown in Fig.~\ref{fig:nspins_chain_results}.
The ground states located near the center of the solution graph are obtained with higher probability.

When the degeneracy is not lifted at first-order perturbation, it is necessary to consider second-order perturbation.
By solving the following eigenvalue equation in second-order perturbation, the ground state $|n^{(0)}\rangle$ corresponding to the smallest eigenvalue $E_n^{(2)}$ is obtained as the final state:
\begin{equation}
E_{n}^{(2)}|n^{(0)}\rangle=P_{2} W P_{2}|n^{(0)}\rangle,
\end{equation}
where $W = V Q \left(E_{n}^{(0)}-H_{0}\right)^{-1} Q V$, $P_{2}$ is the projection operator onto the degenerate ground-state subspace of $E_0^{(1)}$, and $Q=1-P_1$ is the projection operator onto the remaining subspace.

In second-order perturbation, the effective operator $W$ yields matrix elements within the remaining degenerate subspace as
\begin{equation}
\langle g_i|W|g_j\rangle=\sum_{m\notin\mathcal{G}}\frac{\langle g_i|V|m\rangle\langle m|V|g_j\rangle}{E_0^{(0)}-E_m^{(0)}},
\end{equation}
where $\mathcal{G}$ denotes the ground-state manifold of $H_0$.
We define the effective matrix $A^{(2)}:=-P_2WP_2$.
For stoquastic drivers in the computational basis, the off-diagonal elements of $A^{(2)}$ are nonnegative, and $A^{(2)}$ can be interpreted as a "weighted adjacency matrix" on the ground states.
The diagonal element $A^{(2)}_{ii}$ is a sum of contributions from excited states reachable from $|g_i\rangle$ by one application of the driver, weighted by inverse gaps $(E_m^{(0)}-E_0^{(0)})^{-1}$.
The off-diagonal element $A^{(2)}_{ij}$ sums the contributions of intermediate states $|m\rangle$ that connect $|g_i\rangle$ and $|g_j\rangle$ at second order, again weighted by inverse gaps.
Thus, $A^{(2)}$ encodes local energy-barrier information around the ground states.

The solution graph corresponding to $A^{(2)}$ has weighted nodes (self-loops) and edges.
However, to enhance readability and explicitly indicate the required perturbation order, the solution graph presented in this paper shows only the Hamming distance (the number of spin flips required for mutual transition) rather than these weights.

\begin{figure}[tbh]
\centering
\begin{minipage}[c]{0.45\linewidth}
\centering
\subfigure[Three-spin triangle model]{\includegraphics[width=0.6\linewidth]{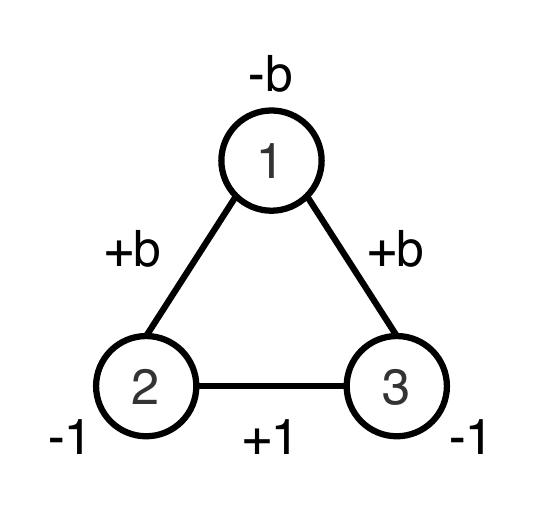}\label{fig:triangle3_model}}
\subfigure[Solution graph]{\includegraphics[width=0.55\linewidth]{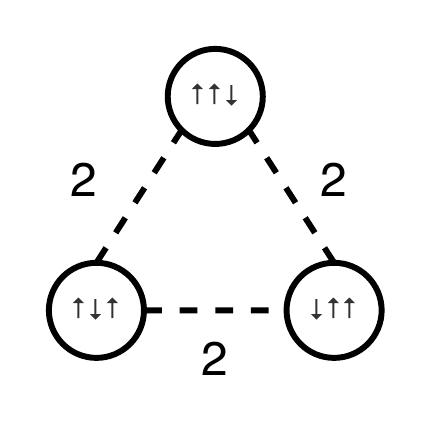}\label{fig:triangle3_solution_graph}}
\end{minipage}
\begin{minipage}[c]{0.45\linewidth}
\centering
\subfigure[Eigenvector centrality and probability vs. relative energy flatness]{\includegraphics[width=\linewidth]{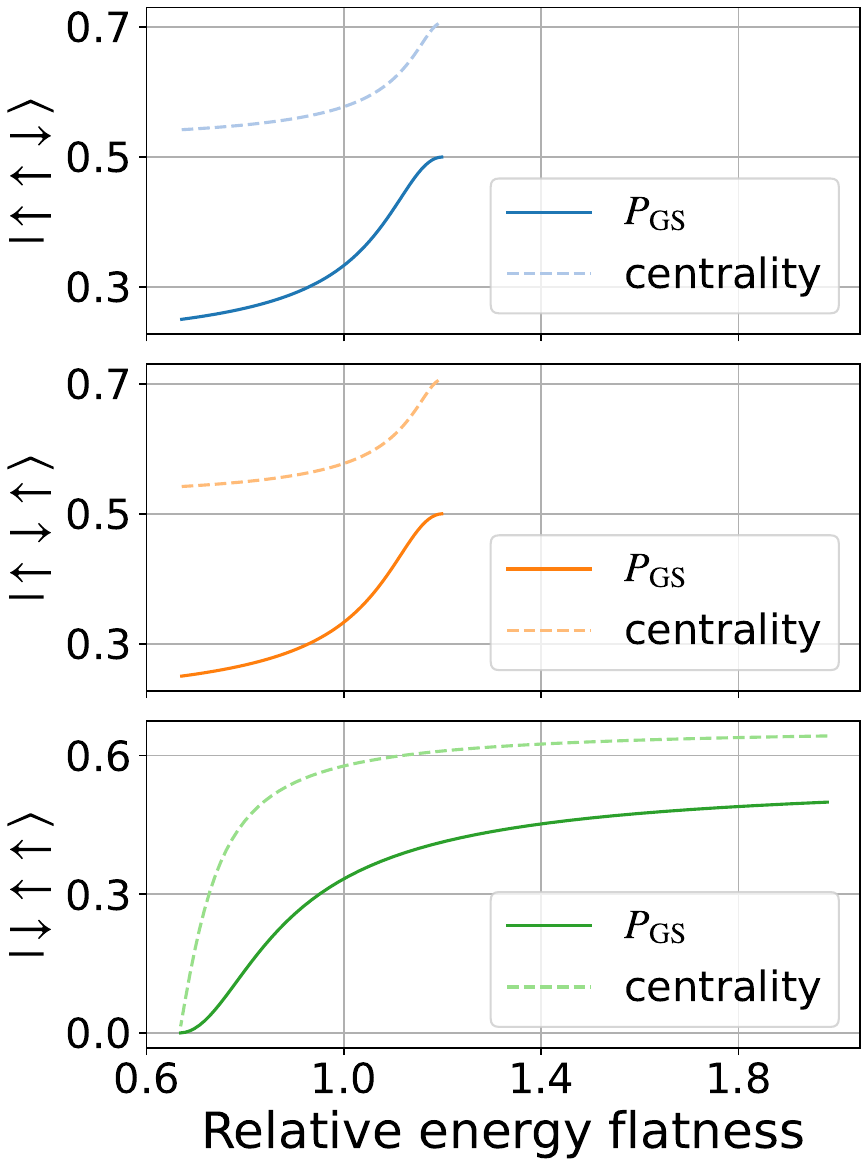}\label{fig:triangle3_centrality}}
\end{minipage}
\caption{%
(a) Three-spin triangle model.
(b) Solution graph in a transverse field.
(c) For each ground state, the probability $P_{\mathrm{GS}}$ (solid) and the eigenvector centrality (dashed) are shown as functions of the relative energy flatness (horizontal axis) when sweeping $b \in (0, 2)$.
We compute centrality using the full weighted matrix $A^{(2)}$, while the weights are omitted from the graph in (b) for clarity.
}
\label{fig:triangle3}
\end{figure}

For a model requiring second-order perturbation, we examine the relationship between eigenvector centrality and the probabilities of ground states.
We consider a triangular lattice model with a parameter $b \in (0,2)$ (Fig.~\ref{fig:triangle3_model}), whose Hamiltonian is given by $H_0=b \sigma_1 \sigma_2+b \sigma_1 \sigma_3+\sigma_2 \sigma_3-b \sigma_1-\sigma_2-\sigma_3$.
This model has three ground states: $|\uparrow \uparrow \downarrow\rangle,|\uparrow \downarrow \uparrow\rangle,|\downarrow \uparrow \uparrow\rangle$.
As shown in Fig.~\ref{fig:triangle3_solution_graph}, these states are mutually connected by two spin flips, so the degeneracy is lifted only at second order when using a transverse-field driver.
In this case, the adjacency matrix is given by
\begin{equation}
A^{(2)} = \begin{pmatrix}
\frac{4-b}{2 b(2-b)} & \frac{1}{b(2-b)} & \frac{1}{b} \\
\frac{1}{b(2-b)} & \frac{4-b}{2 b(2-b)} & \frac{1}{b} \\
\frac{1}{b} & \frac{1}{b} & \frac{4-b}{2 b(2-b)}
\end{pmatrix}.
\label{eq:triangle3_adj_mat}
\end{equation}
In the solution graph shown in Fig.~\ref{fig:triangle3_solution_graph}, the diagonal and off-diagonal elements of $A^{(2)}$ are in fact treated as node and edge weights, respectively, but these are omitted from the figure for clarity.

By varying $b$ and comparing the eigenvector centrality and probability of each ground state, as shown in Fig.~\ref{fig:triangle3_centrality}, we find that their tendencies coincide and are correlated.

Furthermore, the eigenvector centrality derived from $A^{(2)}$, which includes energy differences to surrounding states, can be regarded as representing the flatness of the energy landscape.
We define here the energy flatness as the sum of the reciprocals of energy differences along the paths to the nearest ground states.
The smaller the energy differences between other states, the larger this index becomes, indicating that the energy landscape is flatter.
The energy flatness $\operatorname{EF}_i$ of the $i$-th ground state can be expressed using $A^{(2)}$ as $\operatorname{EF}_i = \sum_{k \neq i} A_{i k}^{(2)}$.
We also define the relative energy flatness of the ground state $i$ as follows:
\begin{equation}
\operatorname{REF}_i = \frac{\operatorname{EF}_i}{\sum_{j \neq i} \operatorname{EF}_j} = \frac{\sum_{k \neq i} A_{i k}^{(2)}}{\sum_{j \neq i} \sum_{k \neq j} A_{j k}^{(2)}}.
\end{equation}
As shown in Eq~\eqref{eq:triangle3_adj_mat}, since all diagonal components are equal in the model, this metric does not include them.

The horizontal axis of Fig.~\ref{fig:triangle3_centrality} represents the relative energy flatness, and the ground states with higher eigenvector centrality have higher flatness and higher probabilities.
Therefore, we can interpret that the parameter $b$ determines the energy landscape, and that the flatness, centrality, and probability change accordingly.

Since the energy landscape around ground states is related to sampling fairness, we expect that problem transformations that modify the landscape will change the fairness.
Energy landscape transformation \cite{fujii2022, fujii2023} is a technique that aims to improve the accuracy of optimal solutions in QA by modifying the landscape through the exchange of biases and interactions in the model.
In the appendix, the results of applying this technique to the toy model are presented, yielding the expected change in sampling fairness.
We also deal with graph minor embedding into actual hardware, which is a similar type of problem transformation, later in this paper.

Even when a higher-order perturbation is required to lift the degeneracy, we expect, as in the second-order case described above, that the energy landscape around ground states can affect their probabilities.
In such cases, the adjacency matrix and solution graph obtained from the perturbative procedure likewise contain information about the energy landscape around the ground states.
As an example of such problems, we present experimental results for the N-Queens problem in the appendix.
Instead of higher-order perturbation analysis, we conducted experiments employing the quantum Monte Carlo method and successfully identified differences in probabilities between fundamental solutions and their variations.
Through simple calculations, we can observe that each fundamental solution corresponds to a different energy landscape.
Based on these findings, we infer that the fairness of ground states remains relevant to the energy landscape even when higher-order perturbations are necessary.

In the experiments so far, we have primarily considered cases where the solution graph consists of a single connected component.
We now consider solution graphs that consist of multiple connected components.
Then, the adjacency matrix is block-diagonalized according to each connected component, allowing us to compute the largest eigenvalue of each block.
If there is a unique largest eigenvalue among these, the final state becomes the eigenstate corresponding to that eigenvalue.
In other words, the ground states belonging to the connected component associated with that eigenvalue dominate in the adiabatic limit.
In contrast, the probability of ground states in other components is suppressed at the leading order, indicating a strong preference.
If two (or more) connected components have exactly the same largest eigenvalue at that order, the effective Hamiltonian remains degenerate, and higher-order corrections (or symmetry-breaking perturbations) are required to determine the final superposition within the enlarged subspace.

\begin{figure}[tbh]
\centering
\subfigure[Solution graphs]{\includegraphics[width=0.8\linewidth]{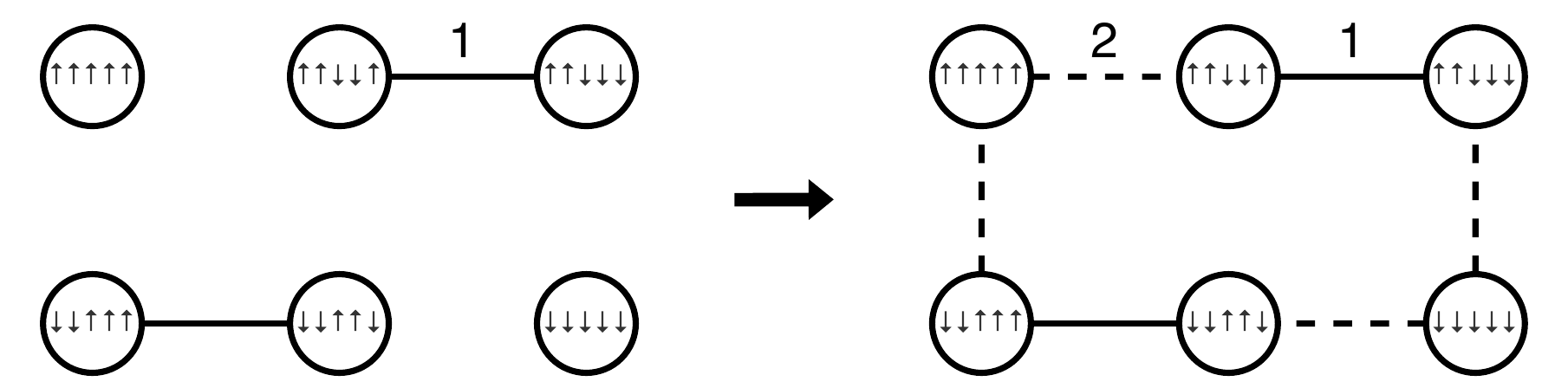}\label{fig:five_spin_matsuda_highorder_solution_graph}}
\par\vspace{1em}
\subfigure[Eigenvector centrality and probability]{\includegraphics[width=0.6\linewidth]{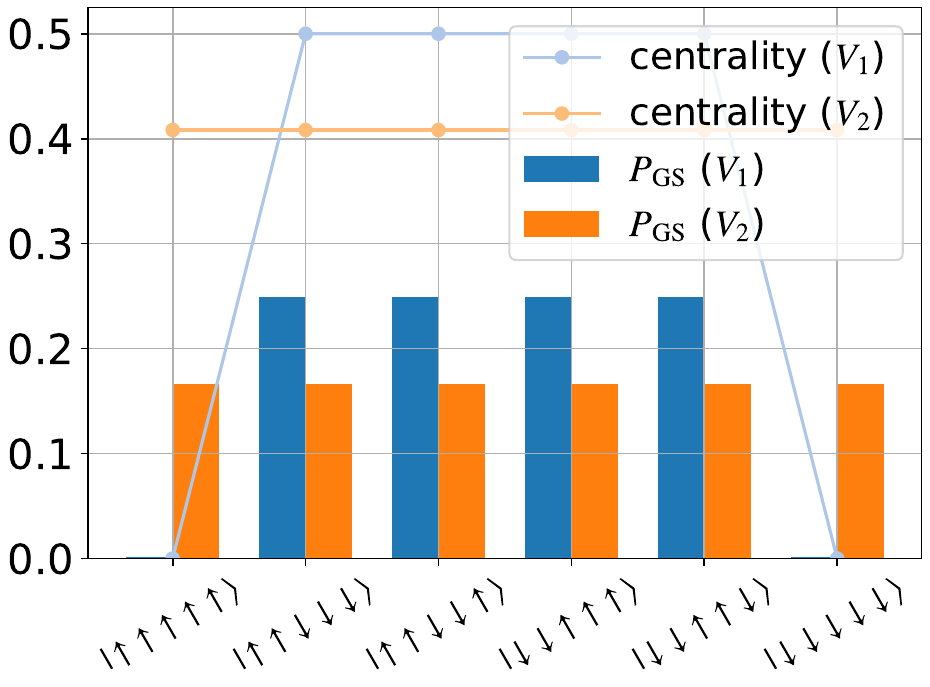}\label{fig:five_spin_matsuda_highorder_centrality}}
\caption{
(a) Solution graphs using a transverse-field $V_1$ (left) and second-order driver $V_2$ (right).
(b) Ground-state probabilities $P_{\mathrm{GS}}$ (bars) compared with the eigenvector centrality (line).
Weights used in the centrality calculations are omitted from the graph drawings for clarity.
}
\label{fig:five_spin_matsuda_highorder}
\end{figure}

\begin{figure}[tbh]
\centering
\subfigure[Five-spin Matsuda models]{\includegraphics[width=0.7\linewidth]{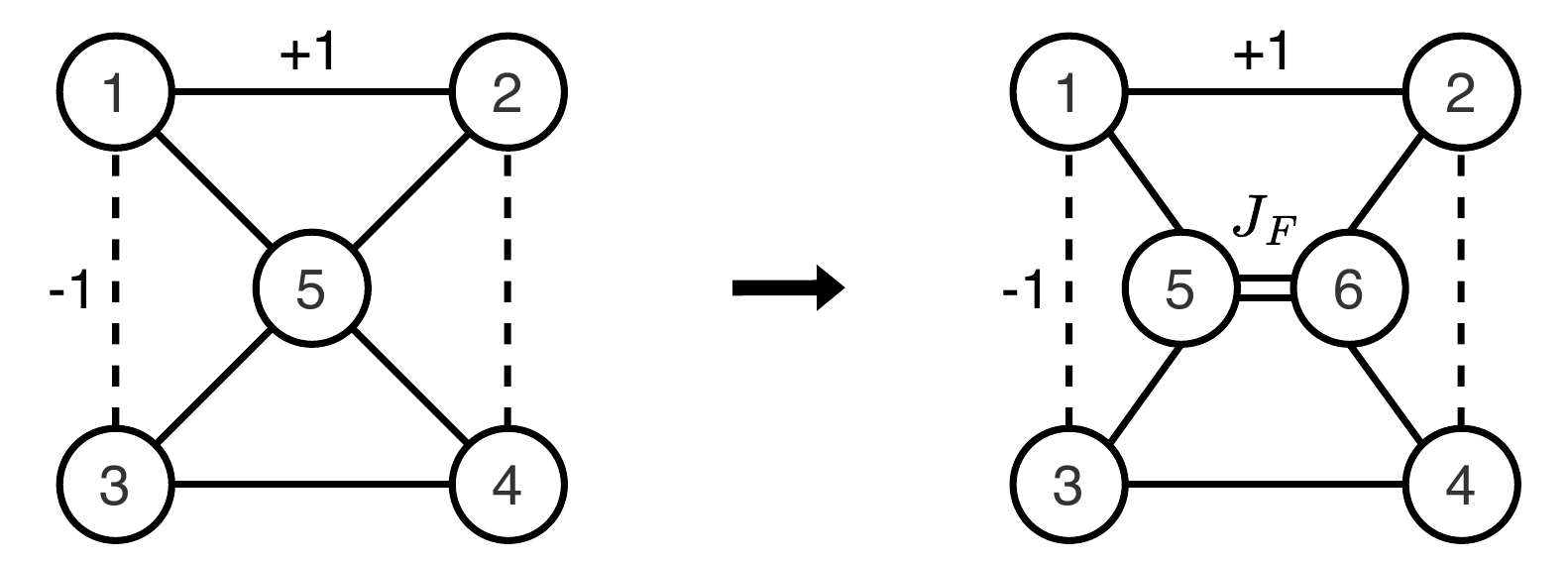}\label{fig:five_spin_matsuda_embedding_model}}
\par\vspace{1em}
\subfigure[Solution graphs]{\includegraphics[width=0.8\linewidth]{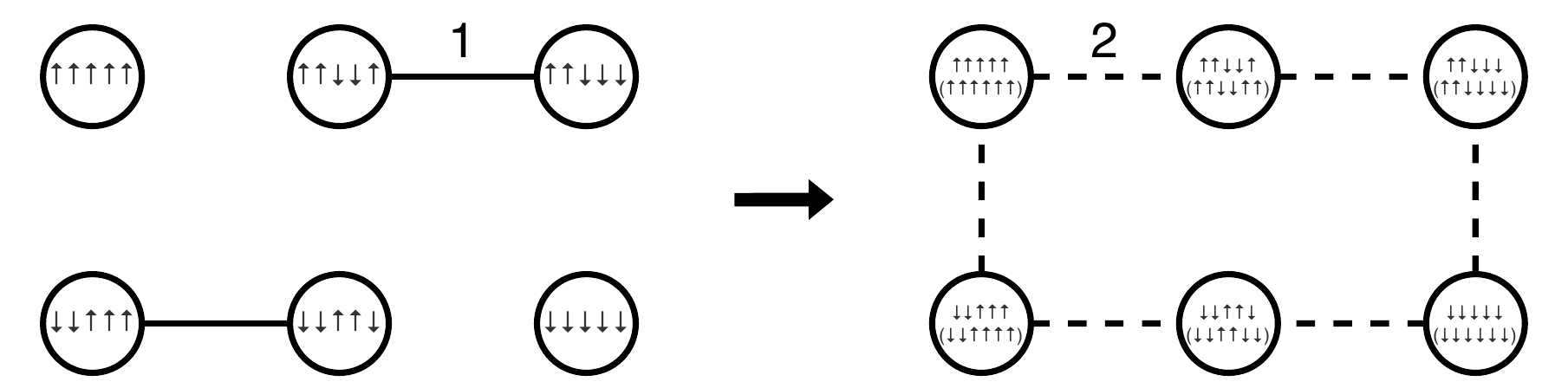}\label{fig:five_spin_matsuda_embedding_solution_graph}}
\par\vspace{1em}
\subfigure[Eigenvector centrality and probability]{\includegraphics[width=0.6\linewidth]{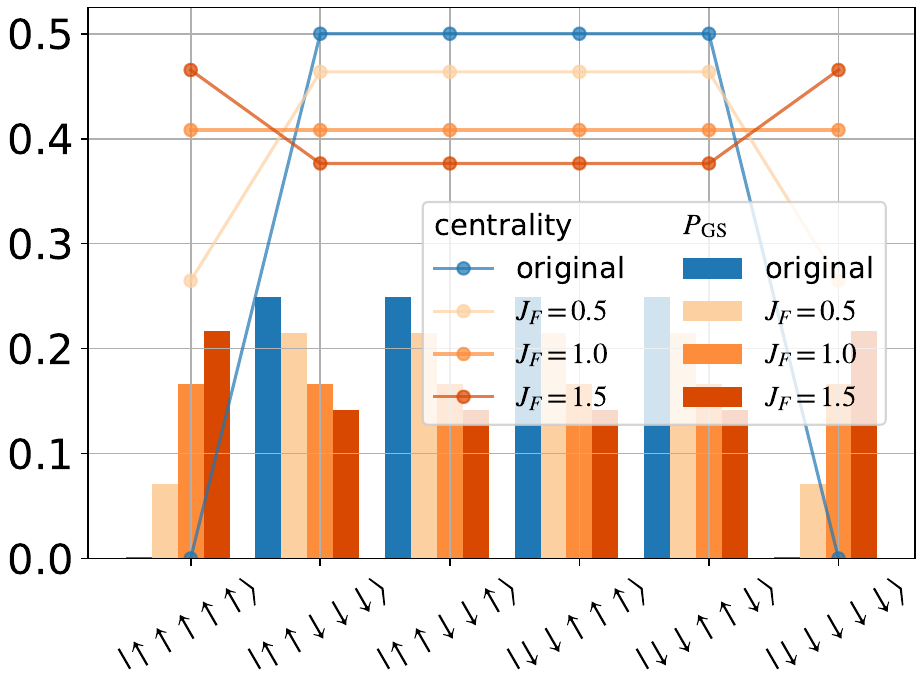}\label{fig:five_spin_matsuda_embedding_centrality}}
\caption{
(a) Five-spin Matsuda models before and after embedding.
(b) Solution graphs in the original and embedded models using a transverse-field driver.
(c) Ground-state probabilities $P_{\mathrm{GS}}$ (bars) compared with the eigenvector centrality (line).
In the embedded model, we set the chain strength $J_F = 0.5, 1.0, 1.5$.
Weights used in the centrality calculations are omitted from the graph drawings for clarity.
}
\label{fig:five_spin_matsuda_embedding}
\end{figure}

We illustrate this behavior using the five-spin Matsuda model \cite{matsuda2009} shown in Fig.~\ref{fig:five_spin_matsuda_embedding_model}.
This model has six ground states $|\uparrow \uparrow \uparrow \uparrow \uparrow\rangle,|\uparrow \uparrow \downarrow \downarrow \uparrow\rangle,|\uparrow \uparrow \downarrow \downarrow \downarrow\rangle,\ldots$ and exhibits spin-inversion and rotational symmetries.
Under the transverse-field driver $V_1 = -\sum_{i=1}^{N} \sigma_i^x$, the solution graph becomes that shown on the left of Fig.~\ref{fig:five_spin_matsuda_highorder_solution_graph}, which contains several connected components of different sizes.
Since the leading eigenvalue is largest in the component with the most nodes, only the ground states $|\uparrow \uparrow \downarrow \downarrow \uparrow\rangle, |\uparrow \uparrow \downarrow \downarrow \downarrow\rangle$ belonging to that component are obtained, as shown in Fig.~\ref{fig:five_spin_matsuda_highorder_centrality}, and the other ground states $|\uparrow \uparrow \uparrow \uparrow \uparrow\rangle$ are never obtained (as well as the spin-reversed states, respectively).
Note that, although the triangular lattice model in Fig.~\ref{fig:triangle3_model} also consists of multiple connected components, the largest eigenvalues of the adjacency matrices in first-order perturbation are all equal, so second-order perturbation is required, and no concentration onto a specific connected component occurs.

Based on the above results, we can summarize that unfair sampling arises when the solution graph has the following two characteristics:
\begin{enumerate}
\item Multiple connected components whose adjacency matrices have different largest eigenvalues: only ground states belonging to the element with the largest eigenvalue are obtained.
\item Differences in eigenvector centrality: states with higher centrality are obtained with higher probability.
\end{enumerate}
Connected components whose adjacency matrices have larger largest eigenvalues tend to contain nodes of higher degree and to be denser.
Let $\bar{d}, \Delta$ be the average and maximum degree of a graph $G$, and $\lambda_1(G)$ the largest eigenvalue of its adjacency matrix, and then $\bar{d} \leq \lambda_1(G) \leq \Delta$ holds as Proposition 3.1.2 \cite{brouwer2012} proves.
Among graphs with the same number of nodes, connected components with larger average or maximum degree therefore have larger largest eigenvalues.
Furthermore, if $G^{\prime}$ is obtained from a connected graph $G$ by deleting a single edge, then $\lambda_1(G^{\prime}) < \lambda_1(G)$ holds as Proposition 3.1.1 \cite{brouwer2012} proves, which means that adding edges to a connected graph strictly increases its largest eigenvalue.
This implies that the largest eigenvalue of the denser component tends to be larger.

By contraposition, we can obtain the following guidelines for reducing sampling bias:
\begin{enumerate}
\item Promote connectivity of the solution graph at the lowest perturbative order.
\item Reduce heterogeneity in the eigenvector centralities.
\end{enumerate}
In the remaining part of this section, we examine previous studies that achieve fair sampling and demonstrate that these cases conform to the above guidelines by interpreting them in terms of solution graphs and centrality.

As a case corresponding to the first guideline, we consider introducing higher-order drivers.
It is known that higher-order drivers can suppress unfairness in sampling \cite{matsuda2009, konz2019}.
In the five-spin Matsuda model described above, using the second-order driver $V_{2} = -\sum_{i} \sigma_{i}^{x} -\sum_{i<j} \sigma_{i}^{x} \sigma_{j}^{x}$ makes all nodes connected to some other node and yields a single connected component (the right of Fig.~\ref{fig:five_spin_matsuda_highorder_solution_graph}).
In particular, the ground states are connected so that their eigenvector centralities all become equal, and the probabilities become exactly equal, as shown in Fig.~\ref{fig:five_spin_matsuda_highorder_centrality}.
However, as indicated by the other study's results \cite{konz2019}, only second-order drivers do not generally achieve complete fairness, and an $N$-th-order driver may be required in the worst case.
Nonetheless, sampling bias can be mitigated by introducing a driver that ensures the solution graph has a single connected component.

Next, we consider graph minor embedding as a case corresponding to the second guideline.
Graph minor embedding maps the problem of interest onto the restricted qubit topology of the quantum annealer.
Because current hardware graphs are not fully connected, it is generally challenging to map logical variables one-to-one to physical qubits.
Instead, multiple physical qubits are associated with a single logical variable.
To be able to read out solutions to the original problem, these qubits are required to take the same state at the end of the anneal, and ferromagnetic couplings are therefore introduced between them (a chain refers to a sequence of qubits connected by such couplings).
Previous work \cite{maruyama2025a} reported that changes in problem structure due to this transformation and changes in the energy landscape caused by the chain strength can affect the fairness of ground states.

As before, we use the five-spin Matsuda model (Fig.~\ref{fig:five_spin_matsuda_embedding_model}).
Before embedding, the states $|\uparrow \uparrow \downarrow \downarrow \uparrow\rangle$ and $|\uparrow \uparrow \downarrow \downarrow \downarrow\rangle$ are obtained with equal probability, whereas the remaining state $|\uparrow \uparrow \uparrow \uparrow \uparrow\rangle$ is never obtained, as shown in Fig.~\ref{fig:five_spin_matsuda_embedding_centrality}.
This preference can be interpreted in terms of eigenvector centrality as mentioned earlier.
Although there are generally many embedding patterns, here we consider the specific embedded model shown on the right side of Fig.~\ref{fig:five_spin_matsuda_embedding_model}, which is compatible with the Pegasus graph of D-Wave Advantage.
In this pattern, the Hamming distance between any pair of ground states is at least two, so all nodes become equally isolated in the solution graph under the transverse-field driver (Fig.~\ref{fig:five_spin_matsuda_embedding_solution_graph}).
A second-order perturbation is therefore required, and we expect that centrality, including contributions from the surrounding energy barriers, is related to the fairness of ground states.
Fig.~\ref{fig:five_spin_matsuda_embedding_centrality} shows the probabilities and eigenvector centralities of each ground state for chain strengths $J_F = 0.5, 1.0, 1.5$.
We observe that the probabilities and centralities exhibit very similar trends.
The results for this embedded model are similar to those for the triangular lattice model (Fig. \ref{fig:triangle3}), indicating that the interaction parameters modify the surrounding energy barriers and centralities, thereby altering the fairness.

From this analysis, graph embedding can serve to "cut" connections between them, whereas higher-order drivers "connect" ground states.
By cutting edges so that each node becomes equally isolated, the bias that makes the probabilities of entire connected components vanish can be removed.
The fairness becomes related to the energy landscape determined by the chain strength.
Graph embedding enables fair sampling, although this depends on the embedding method and its strength, as described in the previous work \cite{maruyama2025a}.
Thus, we see that graph minor embedding is consistent with the guideline of reducing differences in eigenvector centrality.

\section{Discussion}

Using solution graphs composed of ground states and eigenvector centrality, which naturally arise in perturbative analyses of final states in QA, we have shown through several cases that unfair sampling in QA can be interpreted.
As seen in Fig.~\ref{fig:nspins_chain}, when a given driver allows mutual transitions between ground states, states with higher eigenvector centrality are obtained with higher probability.
For problems in which ground states cannot transition directly under the driver, second-order perturbation becomes necessary, and not only adjacency relations but also energy barriers between states become relevant.
Since the flatness of the energy landscape around ground states appears as node and edge weights in the solution graph, eigenvector centrality can be regarded as representing this flatness.
We demonstrated that, similar to the first-order and second-order perturbation ground states, higher eigenvector centrality exhibits higher flatness and probabilities (Fig.~\ref{fig:triangle3}).
Although we used specific toy models in these experiments, the interpretation makes no assumptions about the target Hamiltonian $H_0$, so it can be applied to general problems.

From the interpretation via solution graphs and centrality, we summarize that two approaches are practical for achieving fair sampling: (1) connecting separate connected components in a solution graph, and (2) reducing biases in centrality and flatness.
From Fig.~\ref{fig:five_spin_matsuda_highorder}, we see that introducing higher-order drivers is one way to connect separate components.
Some studies on Quantum Alternating Operator Ansatz \cite{golden2022, pelofske2024a} found that the transverse-field mixer also leads to a strong bias, similar to QA, while the Grover mixer has a theoretical guarantee of fair sampling of degenerate ground states at sufficiently large depth.
On the other hand, implementing higher-order fully connected drivers in the current quantum annealer is a challenging task.
There are also the results showing that simply increasing the order of the driver slightly is insufficient to achieve fairness \cite{konz2019}.
However, our results suggest that to sample ground states equally, it is not necessary for a higher-order driving term to have an order equal to the system size or to be fully connected.
In fact, to construct the solution graph shown on the right of Fig.~\ref{fig:five_spin_matsuda_highorder_solution_graph}, it suffices to use $V_2 = - \sigma_1^x \sigma_2^x - \sigma_3^x \sigma_4^x$ rather than a fully connected driver.
Of course, since we cannot know all ground states of a problem in advance, it is nontrivial to determine which couplings are needed.
Even so, if higher-order drivers are partially implemented in future quantum annealers, the potential for achieving fair sampling should be greater than in the current transverse-field case.

Moreover, embedding problems into hardware graphs can be viewed as a way to cut biased connections between ground states and equalize their flatness depending on the chain strength (Fig.~\ref{fig:five_spin_matsuda_embedding}).
Along this line, there is the possibility of reducing unfairness even on the current quantum annealer.
For example, by repeatedly running heuristic graph embedding and trying various solution graphs and chain strengths, one may be able to realize fair sampling.
Graph embedding is usually regarded as a bottleneck that degrades optimization and sampling accuracy in quantum annealers.
On the other hand, the quantum annealing correction method \cite{vinci2015, bauza2024} enhances optimization performance through the embedding process.
We likewise expect that embedding can be actively exploited to mitigate unfairness among ground states.

Previous works have attempted to explain why central and flat solutions are preferred in QA.
Firstly, the flatness of the landscape is regarded as vital because it helps explain the robustness of solutions in optimization and the generalization performance of models in machine learning \cite{keskar2017}.
Several prior studies have also shown that QA can achieve a state with a flat energy landscape with high probability \cite{ohzeki2018, baldassi2018}.
In this context, flat solutions exhibit high local entropy and hold an advantage in minimizing free energy.
Consequently, it is believed that flat solutions are selected in QA, depending on conditions such as the schedule of quantum fluctuations.
However, there are various possible definitions of flatness, some of which are relatively difficult to interpret intuitively.
One contribution of this study is to relate the flatness of solutions obtained by QA to centrality, providing an interpretation that is easier to grasp intuitively.

If we broaden our view from ground states to low-energy states (approximate solutions), it is known that QA yields diverse low-energy solutions \cite{mohseni2021, zucca2021}.
Sampling diverse solutions is valuable for practical optimization applications, such as enabling better decision making depending on the situation and accelerating exploration in black-box optimization \cite{doi2023, haba2024diverse}.
However, the preference for low-energy solutions is scarcely understood in either theory or experiment.
If a bias exists in low-energy states as well as in the ground state, this intuitively contradicts the concept of diversity.
To accurately assess the validity of existing studies and prospects on the diversity of solutions obtained by QA, it is helpful to understand fairness and diversity in a unified way.
We believe that the interpretation in terms of solution graphs and centrality proposed in this work is a practical approach toward that goal.

\begin{acknowledgments}
This study was supported by the JSPS KAKENHI Grant No. 23H01432.
We received financial supports by programs for bridging the gap between R\&D and IDeal society (Society 5.0) and Generating Economic and social value (BRIDGE) and Cross-ministerial Strategic Innovation Promotion Program (SIP) from the Cabinet Office (No. 23836436).
\end{acknowledgments}

\section*{Author contributions}
N.M. conceived of the presented idea and performed the experiments.
M.O. supervised the findings of this work.
All authors discussed the results and contributed to the preparation of the final manuscript.

\appendix

\section{Energy landscape transformation}

\begin{figure}[tbh]
\centering
\subfigure[Five-spin Matsuda models]{\includegraphics[width=0.9\linewidth]{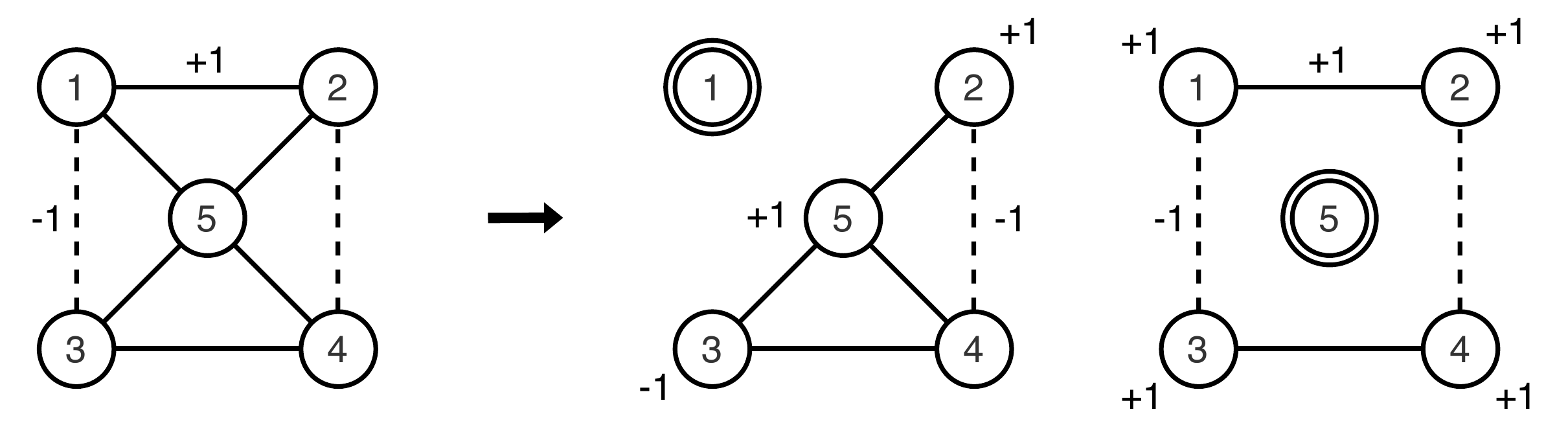}\label{fig:five_spin_matsuda_eltip_model}}
\par\vspace{1em}
\subfigure[Solution graphs]{\includegraphics[width=0.9\linewidth]{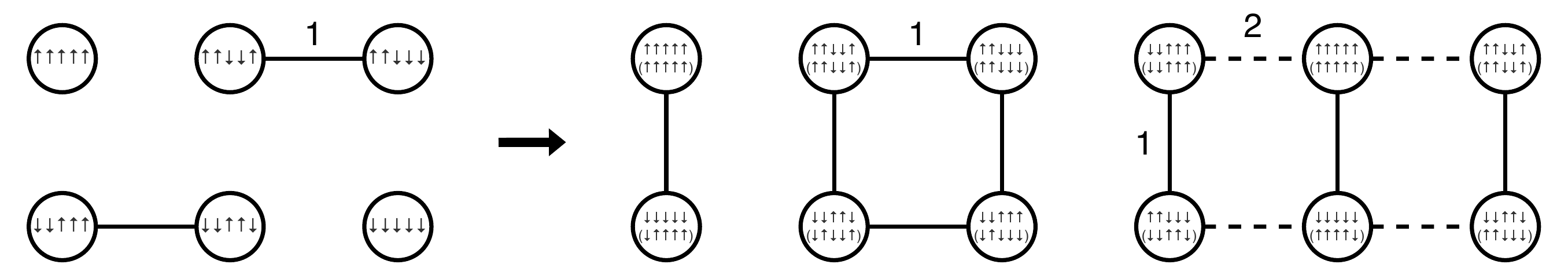}\label{fig:five_spin_matsuda_eltip_solution_graph}}
\par\vspace{1em}
\subfigure[Eigenvector centrality and probability]{\includegraphics[width=0.6\linewidth]{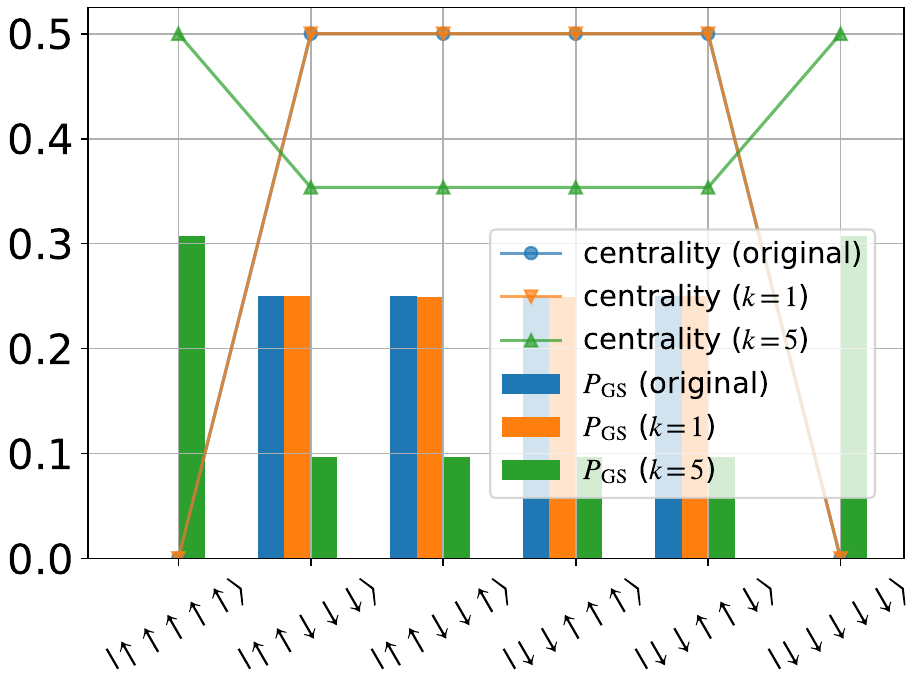}\label{fig:five_spin_matsuda_eltip_centrality}}
\caption{%
(a) Five-spin Matsuda models before and after ELTIP (left: $k=1$, right: $k=5$).
(b) Solution graphs in the original and transformed models using a transverse-field driver.
(c) Ground-state probabilities $P_{\mathrm{GS}}$ (bars) compared with the eigenvector centrality (line).
Weights used in the centrality calculations are omitted from the graph drawings for clarity.
}
\label{fig:five_spin_matsuda_eltip}
\end{figure}

We numerically investigate how energy landscape transformation of Ising problem (ELTIP) affects the fairness of ground states at the final time.
For the five-spin Matsuda model, we use a transverse-field driver and apply ELTIP.
Fig.~\ref{fig:five_spin_matsuda_eltip_model} and Fig.~\ref{fig:five_spin_matsuda_eltip_solution_graph} show the transformed problem and the corresponding solution graphs, respectively.
As one of the ELTIP procedures, we select one spin variable and exchange its local field and its interactions with the other spins to which it is coupled.
In this model, there are two choices: selecting the outer spin ($k=1$) or the central spin ($k=5$).

Fig.~\ref{fig:five_spin_matsuda_eltip_centrality} shows the probabilities and eigenvector centralities of each ground state.
We observe that ELTIP clearly modifies centrality and fairness for $k=1$, although there is no change for $k=5$.
Interpreting this result through solution graphs, we find that the solution graph for $k=1$ has two connected components, and the right component has the larger leading eigenvalue because it has four nodes.
Within that component, all the eigenvector centralities are equal.
Thus, even after ELTIP, the two states $|\uparrow \uparrow \uparrow \uparrow \uparrow\rangle, | \downarrow \downarrow \downarrow \downarrow \downarrow \rangle$ in the left component remain suppressed, while the remaining four states are obtained with equal probabilities.
For $k=5$, second-order perturbation becomes necessary, so in the solution graph, the states that two spin flips can reach are connected by edges.
There are two connected components with identical largest eigenvalues; therefore, neither component is suppressed as a whole.
Within each component, the central state among the three has the most significant eigenvector centrality, and the probabilities differ accordingly.
These observations suggest that changes in the fairness of ground states induced by ELTIP can be consistently interpreted in terms of solution graphs and centrality.

\section{The N-Queens problem requiring higher-order perturbations}

As an example that requires a higher-order perturbation to obtain the final state analytically, we consider the N-Queens problem.
The N-Queens problem is a classical chess puzzle, where $N$ queens are placed on an $N \times N$ chessboard such that no two queens attack each other.
The cost function to be minimized is given by
\begin{equation}
H_0 = \sum_{i=1}^{N}\left(\sum_{j=1}^{N} x_{i j} - 1 \right)^2 + \sum_{j=1}^{N}\left(\sum_{i=1}^{N} x_{i j} - 1 \right)^2 + \sum_{\{D\}} \left(\sum_{(i, j) \in D} x_{ij}\right) \left( \sum_{(i, j) \in D} x_{ij} - 1 \right),
\end{equation}
where $x_{ij} \in \{0,1\}$ is a binary variable that represents whether a queen is placed at row $i$, column $j$.
The set $D$ denotes collections of coordinates along each diagonal direction.
The first and second terms enforce the constraints that each row and each column contains exactly one queen, respectively.
The third term enforces the constraint that along any diagonal direction, the number of queens must be either zero or one.
The N-Queens problem has fundamental solutions, and for each such solution, variant solutions can be obtained by rotations and reflections.
For example, when $N=8$, there exist 12 fundamental solutions and 92 variant solutions in total.

Multiple fundamental solutions appear in the N-Queens problem only when $N \ge 5$, so we focus on instances of this size or larger.
Because rotational and mirror symmetries relate the variant solutions, the Hamming distance between any pair of them is greater than two.
Consequently, using a transverse-field driver, a higher-order perturbation is required to lift the degeneracy.
In addition, since the number of variables is $N^2$, directly solving the Schrödinger equation is infeasible for these sizes.
We therefore simulate QA with a transverse field using the quantum Monte Carlo method (QMC) \cite{king2021}.
We perform 10 independent simulations with different random seeds, using $10^3$ sweeps and $10^5$ samples in each run.

\begin{figure}[tbh]
\centering
\subfigure[$N=5$]{\includegraphics[width=0.45\linewidth]{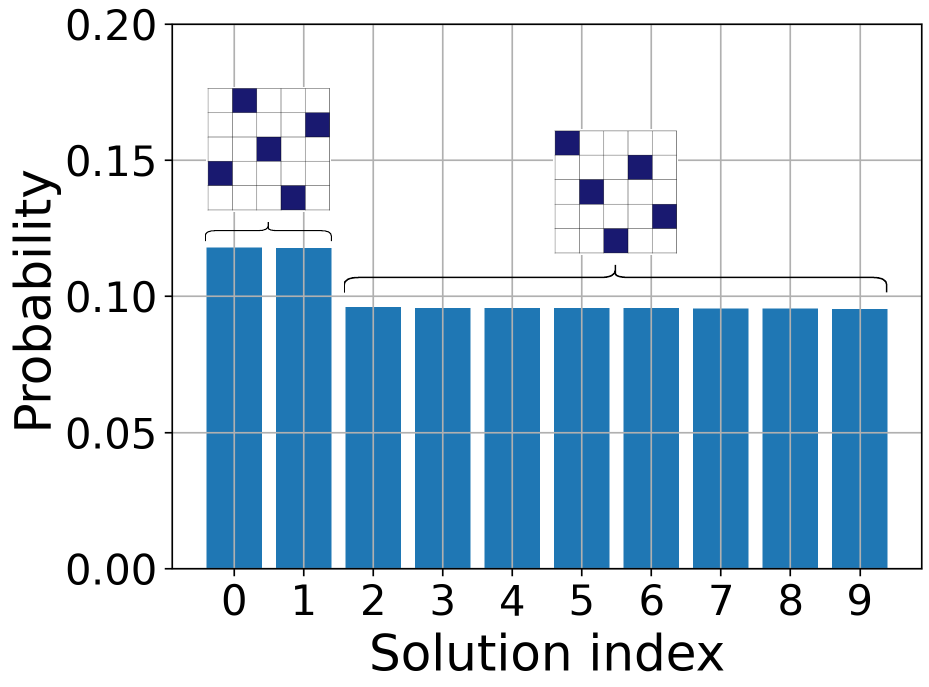}\label{fig:nqueens_probs_N5}}
\hspace{1em}
\subfigure[$N=7$]{\includegraphics[width=0.45\linewidth]{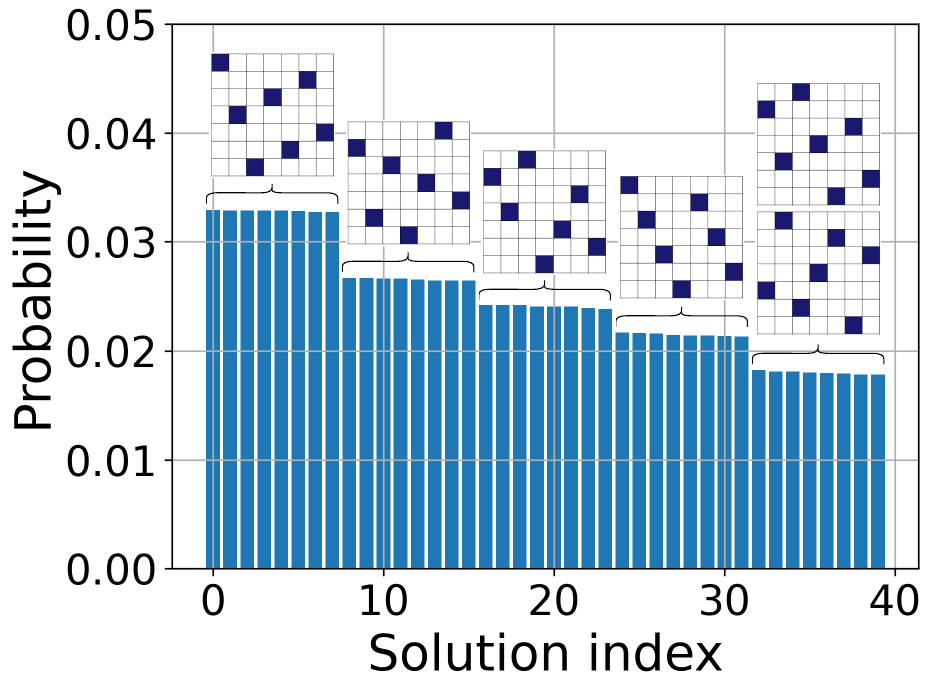}\label{fig:nqueens_probs_N7}}
\caption{
Normalized frequencies (probabilities) of variant solutions obtained by QMC for the N-Queens problem: (a) $N = 5$ and (b) $N = 7$.
Variant solutions are grouped by their associated fundamental solution, shown above each group.
}
\label{fig:nqueens_probs}
\end{figure}

Fig. \ref{fig:nqueens_probs} shows the frequency of each variant solution, where the fundamental solutions are shown above the chart.
For $N=5$, there are two fundamental solutions and 10 variant solutions, whereas for $N=7$, there are six fundamental solutions and 40 variant solutions.
Firstly, we observe that all variant solutions corresponding to the same fundamental solution have identical probabilities.
Secondly, there is a clear bias in the probabilities between different fundamental solutions.
Because higher-order corrections are required for the transverse-field driver, we expect fairness to be related to the surrounding energy landscape, similarly to the second-order case discussed in the main sections.

We next briefly characterize the local energy landscape in the N-Queens problem.
If we define the neighborhood of a solution as the set of states reachable by a single spin flip, there are two possibilities: (1) removing an existing queen and (2) adding a queen to an empty site.
When removing a queen, the energy difference remains the same regardless of which queen is removed and is independent of the solution.
When adding a queen to an empty site, the penalty in the row and column directions is also independent of solutions, because the number of queens changes from one to two.
Along the diagonal directions, however, the number of collisions can be 0, 1, or 2.
We therefore characterize the local energy landscape by the triple $(a,b,c)$, which counts the number of sites with 0, 1, and 2 diagonal conflicts, respectively.
For $N=5$, the two fundamental solutions have $(a,b,c) = (2,12,6)$ and $(4,8,8)$.
The latter solution has more diagonally "safe" sites (larger $a$), so it receives a larger contribution from excited states with smaller gaps in higher-order perturbation, leading to a flatter local energy landscape.
Indeed, as shown in the QMC results (Fig. \ref{fig:nqueens_probs}), the probability of the latter solution is higher than that of the former, in qualitative agreement with this flatness-based argument.
Note that QMC is known to exhibit a uniform bias that hinders faithful simulation of QA in degenerate systems \cite{maruyama2025}.
Thus, while the above results qualitatively capture unfairness, they may quantitatively deviate from those of actual QA.

\bibliography{main}

@article{azinovic2017,
  title = {Assessment of {{Quantum Annealing}} for the {{Construction}} of {{Satisfiability Filters}}},
  author = {Azinovi{\'c}, Marlon and Herr, Daniel and Heim, Bettina and Brown, Ethan and Troyer, Matthias},
  year = 2017,
  month = apr,
  journal = {SciPost Physics},
  volume = {2},
  number = {2},
  pages = {013},
  issn = {2542-4653},
  doi = {10.21468/SciPostPhys.2.2.013},
  urldate = {2021-03-07},
  langid = {english}
}

@article{baldassi2018,
  title = {Efficiency of Quantum vs. Classical Annealing in Nonconvex Learning Problems},
  author = {Baldassi, Carlo and Zecchina, Riccardo},
  year = 2018,
  month = feb,
  journal = {Proceedings of the National Academy of Sciences},
  volume = {115},
  number = {7},
  pages = {1457--1462},
  issn = {0027-8424, 1091-6490},
  doi = {10.1073/pnas.1711456115},
  urldate = {2021-05-15},
  langid = {english}
}

@inproceedings{baptista2018,
  title = {Bayesian {{Optimization}} of {{Combinatorial Structures}}},
  booktitle = {Proceedings of the 35th {{International Conference}} on {{Machine Learning}}},
  author = {Baptista, Ricardo and Poloczek, Matthias},
  year = 2018,
  month = jul,
  pages = {462--471},
  publisher = {PMLR},
  issn = {2640-3498},
  urldate = {2025-10-15},
  langid = {english}
}

@misc{bauza2024,
  title = {Scaling {{Advantage}} in {{Approximate Optimization}} with {{Quantum Annealing}}},
  author = {Bauza, Humberto Munoz and Lidar, Daniel A.},
  year = 2024,
  month = jan,
  number = {arXiv:2401.07184},
  eprint = {2401.07184},
  primaryclass = {cond-mat, physics:quant-ph},
  publisher = {arXiv},
  doi = {10.48550/arXiv.2401.07184},
  urldate = {2024-02-02},
  archiveprefix = {arXiv}
}

@article{boothby2020,
  title = {Next-{{Generation Topology}} of {{D-Wave Quantum Processors}}},
  author = {Boothby, Kelly and Bunyk, Paul and Raymond, Jack and Roy, Aidan},
  year = 2020,
  month = feb,
  journal = {arXiv:2003.00133 [quant-ph]},
  eprint = {2003.00133},
  primaryclass = {quant-ph},
  urldate = {2021-08-05},
  archiveprefix = {arXiv}
}

@incollection{brouwer2012,
  title = {Eigenvalues and {{Eigenvectors}} of {{Graphs}}},
  booktitle = {Spectra of {{Graphs}}},
  author = {Brouwer, Andries E. and Haemers, Willem H.},
  editor = {Brouwer, Andries E. and Haemers, Willem H.},
  year = 2012,
  pages = {33--66},
  publisher = {Springer},
  address = {New York, NY},
  doi = {10.1007/978-1-4614-1939-6_3},
  urldate = {2025-11-29},
  isbn = {978-1-4614-1939-6},
  langid = {english}
}

@article{chancellor2017,
  title = {Modernizing Quantum Annealing Using Local Searches},
  author = {Chancellor, Nicholas},
  year = 2017,
  month = feb,
  journal = {New Journal of Physics},
  volume = {19},
  number = {2},
  pages = {023024},
  publisher = {IOP Publishing},
  issn = {1367-2630},
  doi = {10.1088/1367-2630/aa59c4},
  urldate = {2020-11-16},
  langid = {english}
}

@article{dattani2019,
  title = {Pegasus: {{The}} Second Connectivity Graph for Large-Scale Quantum Annealing Hardware},
  shorttitle = {Pegasus},
  author = {Dattani, Nike and Szalay, Szilard and Chancellor, Nick},
  year = 2019,
  month = jan,
  journal = {arXiv:1901.07636 [quant-ph]},
  eprint = {1901.07636},
  primaryclass = {quant-ph},
  urldate = {2020-02-23},
  archiveprefix = {arXiv}
}

@article{dixit2021a,
  title = {Training {{Restricted Boltzmann Machines With}} a {{D-Wave Quantum Annealer}}},
  author = {Dixit, Vivek and Selvarajan, Raja and Alam, Muhammad A. and Humble, Travis S. and Kais, Sabre},
  year = 2021,
  month = jun,
  journal = {Frontiers in Physics},
  volume = {9},
  publisher = {Frontiers},
  issn = {2296-424X},
  doi = {10.3389/fphy.2021.589626},
  urldate = {2025-08-23},
  langid = {english}
}

@misc{doi2023,
  title = {Exploration of New Chemical Materials Using Black-Box Optimization with the {{D-wave}} Quantum Annealer},
  author = {Doi, Mikiya and Nakao, Yoshihiro and Tanaka, Takuro and Sako, Masami and Ohzeki, Masayuki},
  year = 2023,
  month = dec,
  journal = {arXiv.org},
  doi = {10.3389/fcomp.2023.1286226},
  urldate = {2024-07-23},
  howpublished = {https://arxiv.org/abs/2312.09537v1},
  langid = {english}
}

@article{farhi2001,
  title = {A {{Quantum Adiabatic Evolution Algorithm Applied}} to {{Random Instances}} of an {{NP-Complete Problem}}},
  author = {Farhi, Edward and Goldstone, Jeffrey and Gutmann, Sam and Lapan, Joshua and Lundgren, Andrew and Preda, Daniel},
  year = 2001,
  month = apr,
  journal = {Science},
  volume = {292},
  number = {5516},
  pages = {472--475},
  publisher = {American Association for the Advancement of Science},
  doi = {10.1126/science.1057726},
  urldate = {2023-09-16}
}

@article{feld2019,
  title = {A {{Hybrid Solution Method}} for the {{Capacitated Vehicle Routing Problem Using}} a {{Quantum Annealer}}},
  author = {Feld, Sebastian and Roch, Christoph and Gabor, Thomas and Seidel, Christian and Neukart, Florian and Galter, Isabella and Mauerer, Wolfgang and {Linnhoff-Popien}, Claudia},
  year = 2019,
  month = jun,
  journal = {Frontiers in ICT},
  volume = {6},
  publisher = {Frontiers},
  issn = {2297-198X},
  doi = {10.3389/fict.2019.00013},
  urldate = {2025-11-24},
  langid = {english}
}

@misc{fujii2022,
  title = {Energy Landscape Transformation of {{Ising}} Problem with Invariant Eigenvalues for Quantum Annealing},
  shorttitle = {��量子アニーリングにおける固有値不変のイジング問題のエネルギーランドスケープ変換},
  author = {Fujii, Toru and Komuro, Koshi and Okudaira, Yosuke and Narita, Ryo and Sawada, Masayasu},
  year = 2022,
  month = feb,
  number = {arXiv:2202.05927},
  eprint = {2202.05927},
  primaryclass = {quant-ph},
  publisher = {arXiv},
  doi = {10.48550/arXiv.2202.05927},
  urldate = {2022-09-14},
  archiveprefix = {arXiv}
}

@misc{fujii2023,
  title = {Eigenvalue-Invariant Transformation of {{Ising}} Problem for Anti-Crossing Mitigation in Quantum Annealing},
  author = {Fujii, Toru and Komuro, Koshi and Okudaira, Yosuke and Sawada, Masayasu},
  year = 2023,
  month = jan,
  number = {arXiv:2301.10427},
  eprint = {2301.10427},
  primaryclass = {quant-ph},
  publisher = {arXiv},
  doi = {10.48550/arXiv.2301.10427},
  urldate = {2023-01-28},
  archiveprefix = {arXiv}
}

@article{geman1984,
  title = {Stochastic {{Relaxation}}, {{Gibbs Distributions}}, and the {{Bayesian Restoration}} of {{Images}}},
  author = {Geman, Stuart and Geman, Donald},
  year = 1984,
  month = nov,
  journal = {IEEE Transactions on Pattern Analysis and Machine Intelligence},
  volume = {PAMI-6},
  number = {6},
  pages = {721--741},
  issn = {1939-3539},
  doi = {10.1109/TPAMI.1984.4767596},
  urldate = {2025-08-03}
}

@misc{genin2019,
  title = {Quantum Chemistry on Quantum Annealers},
  author = {Genin, Scott N. and Ryabinkin, Ilya G. and Izmaylov, Artur F.},
  year = 2019,
  month = jan,
  number = {arXiv:1901.04715},
  eprint = {1901.04715},
  primaryclass = {physics},
  publisher = {arXiv},
  doi = {10.48550/arXiv.1901.04715},
  urldate = {2025-08-23},
  archiveprefix = {arXiv}
}

@article{golden2022,
  title = {Fair {{Sampling Error Analysis}} on {{NISQ Devices}}},
  author = {Golden, John and B{\"a}rtschi, Andreas and O'Malley, Daniel and Eidenbenz, Stephan},
  year = 2022,
  month = jun,
  journal = {ACM Transactions on Quantum Computing},
  volume = {3},
  number = {2},
  pages = {1--23},
  issn = {2643-6809, 2643-6817},
  doi = {10.1145/3510857},
  urldate = {2022-05-29},
  langid = {english}
}

@misc{haba2024diverse,
  author       = {Haba, Renichiro and Maki, Daisuke and Tokuhira, Toshiki and Nakamura, Kensaku and Abe, Atsutoshi and Iwakabe, Koichi},
  title        = {Diverse solutions via quantum annealing leads to the discovery of diverse material compositions},
  howpublished = {Presentation at Adiabatic Quantum Computing Conference 2024 (AQC 2024)},
  year         = {2024},
  month        = jun,
  day          = {14},
  note         = {Glasgow, UK}
}

@article{haba2025,
  title = {Routing and Scheduling Optimization for Urban Air Mobility Fleet Management Using Quantum Annealing},
  author = {Haba, Renichiro and Mano, Takuya and Ueda, Ryosuke and Ebe, Genichiro and Takeda, Kohei and Terabe, Masayoshi and Ohzeki, Masayuki},
  year = 2025,
  month = feb,
  journal = {Scientific Reports},
  volume = {15},
  number = {1},
  pages = {4326},
  publisher = {Nature Publishing Group},
  issn = {2045-2322},
  doi = {10.1038/s41598-025-86843-w},
  urldate = {2025-08-02},
  copyright = {2025 The Author(s)},
  langid = {english}
}

@article{johnson2010,
  title = {A Scalable Control System for a Superconducting Adiabatic Quantum Optimization Processor},
  author = {Johnson, M. W. and Bunyk, P. and Maibaum, F. and Tolkacheva, E. and Berkley, A. J. and Chapple, E. M. and Harris, R. and Johansson, J. and Lanting, T. and Perminov, I. and Ladizinsky, E. and Oh, T. and Rose, G.},
  year = 2010,
  month = apr,
  journal = {Superconductor Science and Technology},
  volume = {23},
  number = {6},
  pages = {065004},
  publisher = {IOP Publishing},
  issn = {0953-2048},
  doi = {10.1088/0953-2048/23/6/065004},
  urldate = {2021-03-23},
  langid = {english}
}

@article{kadowaki1998,
  title = {Quantum Annealing in the Transverse {{Ising}} Model},
  author = {Kadowaki, Tadashi and Nishimori, Hidetoshi},
  year = 1998,
  month = nov,
  journal = {Physical Review E},
  volume = {58},
  number = {5},
  pages = {5355--5363},
  publisher = {American Physical Society},
  doi = {10.1103/PhysRevE.58.5355},
  urldate = {2023-09-16}
}

@article{kadowaki2019,
  title = {Experimental and {{Theoretical Study}} of {{Thermodynamic Effects}} in a {{Quantum Annealer}}},
  author = {Kadowaki, Tadashi and Ohzeki, Masayuki},
  year = 2019,
  month = jun,
  journal = {Journal of the Physical Society of Japan},
  volume = {88},
  number = {6},
  pages = {061008},
  publisher = {The Physical Society of Japan},
  issn = {0031-9015},
  doi = {10.7566/JPSJ.88.061008},
  urldate = {2022-12-09}
}

@misc{keskar2017,
  title = {On {{Large-Batch Training}} for {{Deep Learning}}: {{Generalization Gap}} and {{Sharp Minima}}},
  shorttitle = {On {{Large-Batch Training}} for {{Deep Learning}}},
  author = {Keskar, Nitish Shirish and Mudigere, Dheevatsa and Nocedal, Jorge and Smelyanskiy, Mikhail and Tang, Ping Tak Peter},
  year = 2017,
  month = feb,
  number = {arXiv:1609.04836},
  eprint = {1609.04836},
  primaryclass = {cs, math},
  publisher = {arXiv},
  doi = {10.48550/arXiv.1609.04836},
  urldate = {2023-09-13},
  archiveprefix = {arXiv}
}

@article{king2021,
  title = {Scaling Advantage over Path-Integral {{Monte Carlo}} in Quantum Simulation of Geometrically Frustrated Magnets},
  author = {King, Andrew D. and Raymond, Jack and Lanting, Trevor and Isakov, Sergei V. and Mohseni, Masoud and {Poulin-Lamarre}, Gabriel and Ejtemaee, Sara and Bernoudy, William and Ozfidan, Isil and Smirnov, Anatoly Yu and Reis, Mauricio and Altomare, Fabio and Babcock, Michael and Baron, Catia and Berkley, Andrew J. and Boothby, Kelly and Bunyk, Paul I. and Christiani, Holly and Enderud, Colin and Evert, Bram and Harris, Richard and Hoskinson, Emile and Huang, Shuiyuan and Jooya, Kais and Khodabandelou, Ali and Ladizinsky, Nicolas and Li, Ryan and Lott, P. Aaron and MacDonald, Allison J. R. and Marsden, Danica and Marsden, Gaelen and Medina, Teresa and Molavi, Reza and Neufeld, Richard and Norouzpour, Mana and Oh, Travis and Pavlov, Igor and Perminov, Ilya and Prescott, Thomas and Rich, Chris and Sato, Yuki and Sheldan, Benjamin and Sterling, George and Swenson, Loren J. and Tsai, Nicholas and Volkmann, Mark H. and Whittaker, Jed D. and Wilkinson, Warren and Yao, Jason and Neven, Hartmut and Hilton, Jeremy P. and Ladizinsky, Eric and Johnson, Mark W. and Amin, Mohammad H.},
  year = 2021,
  month = feb,
  journal = {Nature Communications},
  volume = {12},
  number = {1},
  pages = {1113},
  publisher = {Nature Publishing Group},
  issn = {2041-1723},
  doi = {10.1038/s41467-021-20901-5},
  urldate = {2025-11-09},
  copyright = {2021 The Author(s)},
  langid = {english}
}

@article{king2022,
  title = {Coherent Quantum Annealing in a Programmable 2000-Qubit {{Ising}} Chain},
  author = {King, Andrew D. and Suzuki, Sei and Raymond, Jack and Zucca, Alex and Lanting, Trevor and Altomare, Fabio and Berkley, Andrew J. and Ejtemaee, Sara and Hoskinson, Emile and Huang, Shuiyuan and Ladizinsky, Eric and MacDonald, Allison and Marsden, Gaelen and Oh, Travis and {Poulin-Lamarre}, Gabriel and Reis, Mauricio and Rich, Chris and Sato, Yuki and Whittaker, Jed D. and Yao, Jason and Harris, Richard and Lidar, Daniel A. and Nishimori, Hidetoshi and Amin, Mohammad H.},
  year = 2022,
  month = nov,
  journal = {Nature Physics},
  volume = {18},
  number = {11},
  eprint = {2202.05847},
  primaryclass = {quant-ph},
  pages = {1324--1328},
  issn = {1745-2473, 1745-2481},
  doi = {10.1038/s41567-022-01741-6},
  urldate = {2025-08-23},
  archiveprefix = {arXiv}
}

@article{kitai2020,
  title = {Designing Metamaterials with Quantum Annealing and Factorization Machines},
  author = {Kitai, Koki and Guo, Jiang and Ju, Shenghong and Tanaka, Shu and Tsuda, Koji and Shiomi, Junichiro and Tamura, Ryo},
  year = 2020,
  month = mar,
  journal = {Physical Review Research},
  volume = {2},
  number = {1},
  pages = {013319},
  publisher = {American Physical Society},
  doi = {10.1103/PhysRevResearch.2.013319},
  urldate = {2025-11-24}
}

@article{konz2019,
  title = {Uncertain Fate of Fair Sampling in Quantum Annealing},
  author = {K{\"o}nz, Mario S. and Mazzola, Guglielmo and Ochoa, Andrew J. and Katzgraber, Helmut G. and Troyer, Matthias},
  year = 2019,
  month = sep,
  journal = {Physical Review A},
  volume = {100},
  number = {3},
  pages = {030303},
  publisher = {American Physical Society},
  doi = {10.1103/PhysRevA.100.030303},
  urldate = {2021-03-16}
}

@article{koshikawa2021,
  title = {Benchmark Test of {{Black-box}} Optimization Using {{D-Wave}} Quantum Annealer},
  author = {Koshikawa, Ami S. and Ohzeki, Masayuki and Kadowaki, Tadashi and Tanaka, Kazuyuki},
  year = 2021,
  month = jun,
  journal = {Journal of the Physical Society of Japan},
  volume = {90},
  number = {6},
  eprint = {2103.12320},
  pages = {064001},
  issn = {0031-9015, 1347-4073},
  doi = {10.7566/JPSJ.90.064001},
  urldate = {2021-10-07},
  archiveprefix = {arXiv}
}

@misc{kumar2020,
  title = {Achieving Fair Sampling in Quantum Annealing},
  author = {Kumar, Vaibhaw and Tomlin, Casey and Nehrkorn, Curt and O'Malley, Daniel and Dulny III, Joseph},
  year = 2020,
  month = sep,
  number = {arXiv:2007.08487},
  eprint = {2007.08487},
  primaryclass = {quant-ph},
  publisher = {arXiv},
  doi = {10.48550/arXiv.2007.08487},
  urldate = {2024-08-14},
  archiveprefix = {arXiv},
  langid = {american}
}

@article{mandra2017,
  title = {Exponentially {{Biased Ground-State Sampling}} of {{Quantum Annealing Machines}} with {{Transverse-Field Driving Hamiltonians}}},
  author = {Mandr{\`a}, Salvatore and Zhu, Zheng and Katzgraber, Helmut G.},
  year = 2017,
  month = feb,
  journal = {Physical Review Letters},
  volume = {118},
  number = {7},
  pages = {070502},
  publisher = {American Physical Society},
  doi = {10.1103/PhysRevLett.118.070502},
  urldate = {2021-01-27}
}

@misc{maruyama2025,
  title = {Uniformity {{Bias}} in {{Ground-State Sampling Induced}} by {{Replica Alignment}} in {{Quantum Monte Carlo}} for {{Quantum Annealing}}},
  author = {Maruyama, Naoki and Ohzeki, Masayuki and Tanaka, Kazuyuki},
  year = 2025,
  month = oct,
  number = {arXiv:2510.10566},
  eprint = {2510.10566},
  primaryclass = {quant-ph},
  publisher = {arXiv},
  doi = {10.48550/arXiv.2510.10566},
  urldate = {2025-10-15},
  archiveprefix = {arXiv},
  langid = {american}
}

@misc{maruyama2025a,
  title = {Graph Minor Embedding Can Affect Sampling Degenerate Ground States Using Quantum Annealing},
  author = {Maruyama, Naoki and Ohzeki, Masayuki and Tanaka, Kazuyuki},
  year = 2025,
  month = nov,
  number = {arXiv:2110.10930},
  eprint = {2110.10930},
  primaryclass = {quant-ph},
  publisher = {arXiv},
  doi = {10.48550/arXiv.2110.10930},
  urldate = {2025-11-30},
  archiveprefix = {arXiv}
}

@article{matsuda2009,
  title = {Ground-State Statistics from Annealing Algorithms: Quantum versus Classical Approaches},
  shorttitle = {Ground-State Statistics from Annealing Algorithms},
  author = {Matsuda, Yoshiki and Nishimori, Hidetoshi and Katzgraber, Helmut G.},
  year = 2009,
  month = jul,
  journal = {New Journal of Physics},
  volume = {11},
  number = {7},
  pages = {073021},
  publisher = {IOP Publishing},
  issn = {1367-2630},
  doi = {10.1088/1367-2630/11/7/073021},
  urldate = {2021-01-27},
  langid = {english}
}

@misc{mohseni2021,
  title = {Diversity Measure for Discrete Optimization: {{Sampling}} Rare Solutions via Algorithmic Quantum Annealing},
  shorttitle = {Diversity Measure for Discrete Optimization},
  author = {Mohseni, Masoud and Rams, Marek M. and Isakov, Sergei V. and Eppens, Daniel and Pielawa, Susanne and Strumpfer, Johan and Boixo, Sergio and Neven, Hartmut},
  year = 2021,
  month = oct,
  number = {arXiv:2110.10560},
  eprint = {2110.10560},
  primaryclass = {cond-mat, physics:quant-ph},
  publisher = {arXiv},
  doi = {10.48550/arXiv.2110.10560},
  urldate = {2022-09-14},
  archiveprefix = {arXiv}
}

@article{morita2008,
  title = {Mathematical Foundation of Quantum Annealing},
  author = {Morita, Satoshi and Nishimori, Hidetoshi},
  year = 2008,
  month = dec,
  journal = {Journal of Mathematical Physics},
  volume = {49},
  number = {12},
  pages = {125210},
  issn = {0022-2488},
  doi = {10.1063/1.2995837},
  urldate = {2023-09-16}
}

@article{neukart2017,
  title = {Traffic {{Flow Optimization Using}} a {{Quantum Annealer}}},
  author = {Neukart, Florian and Compostella, Gabriele and Seidel, Christian and {von Dollen}, David and Yarkoni, Sheir and Parney, Bob},
  year = 2017,
  month = dec,
  journal = {Frontiers in ICT},
  volume = {4},
  publisher = {Frontiers},
  issn = {2297-198X},
  doi = {10.3389/fict.2017.00029},
  urldate = {2025-08-01},
  langid = {english}
}

@inproceedings{neven2012,
  title = {{{QBoost}}: {{Large Scale Classifier Training withAdiabatic Quantum Optimization}}},
  shorttitle = {{{QBoost}}},
  booktitle = {Proceedings of the {{Asian Conference}} on {{Machine Learning}}},
  author = {Neven, Hartmut and Denchev, Vasil S. and Rose, Geordie and Macready, William G.},
  year = 2012,
  month = nov,
  pages = {333--348},
  publisher = {PMLR},
  issn = {1938-7228},
  urldate = {2025-08-23},
  langid = {english}
}

@article{ohzeki2018,
  title = {Optimization of Neural Networks via Finite-Value Quantum Fluctuations},
  shorttitle = {��有限値量子ゆらぎを利用したニューラルネットワークの最適化},
  author = {Ohzeki, Masayuki and Okada, Shuntaro and Terabe, Masayoshi and Taguchi, Shinichiro},
  year = 2018,
  month = jul,
  journal = {Scientific Reports},
  volume = {8},
  number = {1},
  pages = {9950},
  issn = {2045-2322},
  doi = {10.1038/s41598-018-28212-4},
  urldate = {2023-01-26},
  langid = {english}
}

@article{otsuka2025a,
  title = {Filtering out Mislabeled Training Instances Using Black-Box Optimization and Quantum Annealing},
  author = {Otsuka, Makoto and Kodama, Kento and Morita, Keisuke and Ohzeki, Masayuki},
  year = 2025,
  month = oct,
  journal = {Scientific Reports},
  volume = {15},
  number = {1},
  pages = {37892},
  publisher = {Nature Publishing Group},
  issn = {2045-2322},
  doi = {10.1038/s41598-025-21686-z},
  urldate = {2025-11-24},
  copyright = {2025 The Author(s)},
  langid = {english}
}

@misc{pelofske2023,
  title = {Comparing {{Three Generations}} of {{D-Wave Quantum Annealers}} for {{Minor Embedded Combinatorial Optimization Problems}}},
  author = {Pelofske, Elijah},
  year = 2023,
  month = jan,
  number = {arXiv:2301.03009},
  eprint = {2301.03009},
  primaryclass = {quant-ph},
  publisher = {arXiv},
  doi = {10.48550/arXiv.2301.03009},
  urldate = {2023-02-25},
  archiveprefix = {arXiv}
}

@misc{pelofske2024a,
  title = {Biased {{Degenerate Ground-State Sampling}} of {{Small Ising Models}} with {{Converged QAOA}}},
  author = {Pelofske, Elijah},
  year = 2024,
  month = nov,
  number = {arXiv:2411.05294},
  eprint = {2411.05294},
  primaryclass = {quant-ph},
  publisher = {arXiv},
  doi = {10.48550/arXiv.2411.05294},
  urldate = {2024-12-07},
  archiveprefix = {arXiv}
}

@misc{sato2021,
  title = {Assessment of Image Generation by Quantum Annealer},
  author = {Sato, Takehito and Ohzeki, Masayuki and Tanaka, Kazuyuki},
  year = 2021,
  month = mar,
  journal = {arXiv.org},
  urldate = {2024-07-23},
  howpublished = {https://arxiv.org/abs/2103.08373v1},
  langid = {english}
}

@misc{sawamura2025,
  title = {Quantum-Classical Hybrid Algorithm Using Quantum Annealing for Multi-Objective Job Shop Scheduling},
  author = {Sawamura, Kenta and Araki, Kensuke and Maruyama, Naoki and Haba, Renichiro and Ohzeki, Masayuki},
  year = 2025,
  month = nov,
  number = {arXiv:2511.03257},
  eprint = {2511.03257},
  primaryclass = {quant-ph},
  publisher = {arXiv},
  doi = {10.48550/arXiv.2511.03257},
  urldate = {2025-11-24},
  archiveprefix = {arXiv}
}

@article{shikanai2025,
  title = {Quadratic {{Unconstrained Binary Formulation}} for {{Traffic Signal Optimization}} on {{Real-World Maps}}},
  author = {Shikanai, Reo and Ohzeki, Masayuki and Tanaka, Kazuyuki},
  year = 2025,
  month = feb,
  journal = {Journal of the Physical Society of Japan},
  volume = {94},
  number = {2},
  pages = {024001},
  publisher = {The Physical Society of Japan},
  issn = {0031-9015},
  doi = {10.7566/JPSJ.94.024001},
  urldate = {2025-08-23}
}

@article{sieberer2018,
  title = {Programmable Superpositions of {{Ising}} Configurations},
  author = {Sieberer, Lukas M. and Lechner, Wolfgang},
  year = 2018,
  month = may,
  journal = {Physical Review A},
  volume = {97},
  number = {5},
  pages = {052329},
  publisher = {American Physical Society},
  doi = {10.1103/PhysRevA.97.052329},
  urldate = {2020-11-16},
  langid = {american}
}

@article{somma2007,
  title = {Quantum {{Approach}} to {{Classical Statistical Mechanics}}},
  author = {Somma, R. D. and Batista, C. D. and Ortiz, G.},
  year = 2007,
  month = jul,
  journal = {Physical Review Letters},
  volume = {99},
  number = {3},
  pages = {030603},
  publisher = {American Physical Society},
  doi = {10.1103/PhysRevLett.99.030603},
  urldate = {2021-02-02}
}

@misc{streif2019,
  title = {Solving {{Quantum Chemistry Problems}} with a {{D-Wave Quantum Annealer}}},
  author = {Streif, Michael and Neukart, Florian and Leib, Martin},
  year = 2019,
  month = mar,
  number = {arXiv:1811.05256},
  eprint = {1811.05256},
  primaryclass = {quant-ph},
  publisher = {arXiv},
  doi = {10.48550/arXiv.1811.05256},
  urldate = {2025-08-23},
  archiveprefix = {arXiv}
}

@article{venturelli2016,
  title = {Quantum {{Annealing Implementation}} of {{Job-Shop Scheduling}}},
  author = {Venturelli, Davide and Marchand, Dominic J. J. and Rojo, Galo},
  year = 2016,
  month = oct,
  journal = {arXiv:1506.08479 [quant-ph]},
  eprint = {1506.08479},
  primaryclass = {quant-ph},
  urldate = {2021-04-05},
  archiveprefix = {arXiv}
}

@article{vinci2015,
  title = {Quantum Annealing Correction with Minor Embedding},
  author = {Vinci, Walter and Albash, Tameem and {Paz-Silva}, Gerardo and Hen, Itay and Lidar, Daniel A.},
  year = 2015,
  month = oct,
  journal = {Physical Review A},
  volume = {92},
  number = {4},
  pages = {042310},
  publisher = {American Physical Society},
  doi = {10.1103/PhysRevA.92.042310},
  urldate = {2024-02-02}
}

@article{weaver2012,
  title = {Satisfiability-Based {{Set Membership Filters}}},
  author = {Weaver, Sean A. and Ray, Katrina J. and Marek, Victor W. and Mayer, Andrew J. and Walker, Alden K.},
  year = 2012,
  month = jan,
  journal = {Journal on Satisfiability, Boolean Modeling and Computation},
  volume = {8},
  number = {3-4},
  pages = {129--148},
  publisher = {IOS Press},
  doi = {10.3233/SAT190095},
  urldate = {2021-03-07},
  langid = {english}
}

@article{yamamoto2020,
  title = {Fair {{Sampling}} by {{Simulated Annealing}} on {{Quantum Annealer}}},
  author = {Yamamoto, Masayuki and Ohzeki, Masayuki and Tanaka, Kazuyuki},
  year = 2020,
  month = jan,
  journal = {Journal of the Physical Society of Japan},
  volume = {89},
  number = {2},
  pages = {025002},
  publisher = {The Physical Society of Japan},
  issn = {0031-9015},
  doi = {10.7566/JPSJ.89.025002},
  urldate = {2021-03-16}
}

@article{zhu2019,
  title = {Fair Sampling of Ground-State Configurations of Binary Optimization Problems},
  author = {Zhu, Zheng and Ochoa, Andrew J. and Katzgraber, Helmut G.},
  year = 2019,
  month = jun,
  journal = {Physical Review E},
  volume = {99},
  number = {6},
  pages = {063314},
  publisher = {American Physical Society},
  doi = {10.1103/PhysRevE.99.063314},
  urldate = {2022-09-22}
}

@misc{zucca2021,
  title = {Diversity Metric for Evaluation of Quantum Annealing},
  author = {Zucca, Alex and Sadeghi, Hossein and Mohseni, Masoud and Amin, Mohammad H.},
  year = 2021,
  month = oct,
  number = {arXiv:2110.10196},
  eprint = {2110.10196},
  primaryclass = {quant-ph},
  publisher = {arXiv},
  doi = {10.48550/arXiv.2110.10196},
  urldate = {2022-09-14},
  archiveprefix = {arXiv}
}

\end{document}